# Low Resistance *P*-Type Contacts to Monolayer $WSe_2$ through Chlorinated Solvent Doping

Lauren Hoang[1], Robert K.A. Bennett[1], Anh Tuan Hoang[2], Tara Pena[1], Zhepeng Zhang[2], Marisa Hocking[2], Ashley P. Saunders[3], Fang Liu[3], Eric Pop[1,2,4], and Andrew J. Mannix[2,*]

[1]*Dept. of Electrical Engineering, Stanford University, Stanford, CA 94305, U.S.A.*

[2]*Dept. of Materials Science and Engineering, Stanford University, Stanford, CA 94305, U.S.A.*

[3]*Dept. of Chemistry, Stanford University, Stanford, CA 94305, U.S.A.*

[4]*Dept. of Applied Physics, Stanford University, Stanford, CA 94305, U.S.A.*

**Tungsten diselenide ($WSe_2$) is a promising *p*-type semiconductor limited by high contact resistance ($R_C$) and the lack of a reliable doping strategy. Here, we demonstrate that exposing $WSe_2$ to chloroform provides simple and stable *p*-type doping. In monolayer $WSe_2$ transistors with Pd contacts, chloroform increases the maximum hole current by over 100× (>200 µA/µm), reduces $R_C$ to ~2.5 kΩ·µm, and retains an on/off ratio of $10^{10}$ at room temperature. These improvements persist for over 8 months, survive annealing above 150 °C, and remain effective down to 10 K, enabling a cryogenic $R_C$ of ~1 kΩ·µm. Density functional theory indicates that chloroform strongly physisorbs to $WSe_2$, inducing hole doping with minimal impact on the electronic states between the valence band and conduction band edges. Auger electron spectroscopy and atomic force microscopy reveal that chloroform intercalates at the $WSe_2$ interface with the gate oxide, contributing to doping stability and mitigating interfacial dielectric disorder. This robust, scalable approach enables high-yield $WSe_2$ transistors with good *p*-type performance.**

Two-dimensional (2D) semiconductors, particularly transition metal dichalcogenides (TMDs), are promising candidates for next-generation, high-density, complementary-metal-oxide-semiconductor (CMOS)[1,2] and low temperature electronics. However, the large contact resistance ($R_C$) often observed in nanoscale TMD devices poses a significant obstacle to device performance, limiting the on-state drain current, $I_D$, needed for practical circuit applications. Both *n*-type and *p*-type transistors are critical for low-power CMOS[3], but progress on minimizing $R_C$ has largely been limited to *n*-type devices[4,5]. Developing scalable, low-$R_C$ *p*-type contacts for 2D transistors remains a critical challenge. Additionally, $R_C$ typically increases further at low temperatures, impeding other fundamental charge transport studies[6].

Various strategies have been explored to reduce $R_C$ to *p*-type $WSe_2$ transistors, including transferred metal[7] or semimetal[8,9] contacts. However, metal contacts typically form large Schottky barriers at the metal-2D semiconductor interface (preventing low $R_C$), and semimetal contacts have yet to experimentally demonstrate superior performance for *p*-type devices. An alternative is to lower $R_C$ by implementing stable *p*-type doping near the contacts. Substitutional doping with electron acceptors (e.g. V[10], Nb[11]) is stable due to covalent chemical bonding but is likely to require complex fabrication with multiple material growth steps. In comparison, *p*-type surface charge transfer doping (SCTD) withdraws electrons from the 2D channel using higher electronegativity capping or adsorbate layers with work function values below the Fermi level of the $WSe_2$, such as transition metal oxides ($MoO_x$[12] and $WO_x$[13,14]), $NO_x$[15,16], and halide compounds ($HAuCl_4$[17], $AuCl_3$[18], $RuCl_3$[19,20], $PtCl_4$[21]) (**Figure 1a**). SCTD typically preserves the host lattice and has the potential to introduce fewer scattering centers[22]. However, the temporal and thermal stability of these methods remain unclear due to the high chemical reactivity or low thermal stability of the reagents involved[23]. Furthermore, there is little consensus on the mechanism of halide-based doping, with some studies suggesting reactions in which Cl atoms substitute and passivate chalcogen vacancies[24–26], while others propose molecular physisorption[18,27] or intercalation[28].

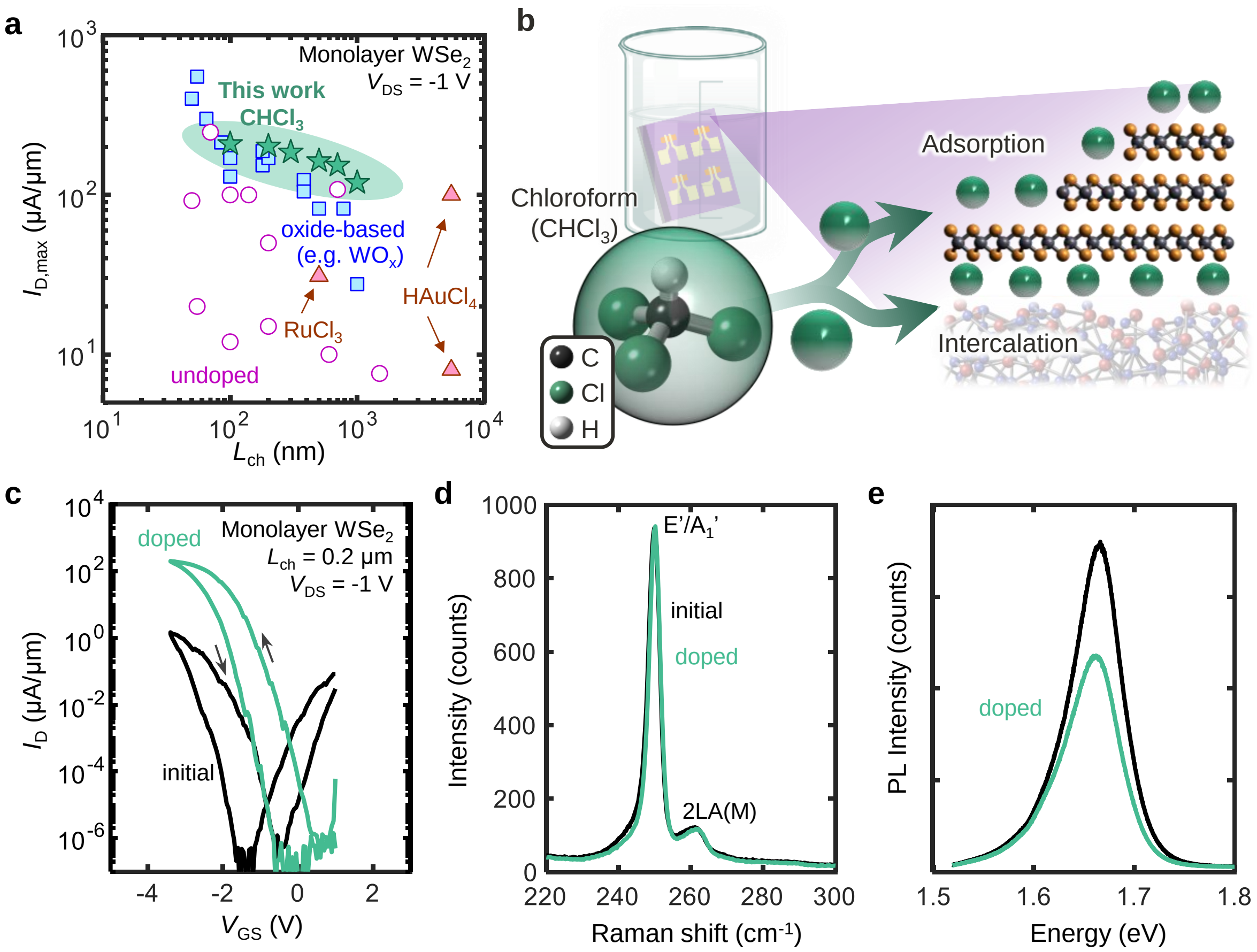

**Figure 1**. ***p*-type Doping of Monolayer $WSe_2$ using Chloroform.** **a**, Benchmarking maximum *p*-type current $I_{D,max}$ vs. monolayer $WSe_2$ channel length ($L_{ch}$) at $V_{DS}$ = -1 V at room temperature, using various contact metals and doping strategies. Circles mark results with no intentional doping, squares denote oxide-based doping ($MoO_x$, $WO_x$, $NO_x$), and triangles label halide-based doping. Our results with chloroform doping (stars) achieve among the highest hole currents to date for monolayer $WSe_2$. **b**, Schematic of chloroform-doped $WSe_2$, illustrating the process and possible adsorption pathways. After fabrication, devices are left in chloroform overnight. **c**, Measured $I_D$ vs. $V_{GS}$ for monolayer $WSe_2$ device before (black line) and after (green line) chloroform doping, reaching hole current of 203 μA/μm. Forward and backward sweeps are shown, revealing some counterclockwise hysteresis. **d**, Raman spectra before and after chloroform doping of monolayer $WSe_2$. **e**, Photoluminescence (PL) spectra of monolayer $WSe_2$ before and after chloroform doping.

Solvent exposure can also unintentionally dope TMDs. For example, $MoS_2$ and $WSe_2$ are *n*-doped by exposure to the low-electronegativity solvent acetone during removal of electron-beam (e-beam) lithography resists like poly(methyl methacrylate) (PMMA)[29,30]. Conversely, the high-electronegativity solvent chloroform ($CHCl_3$) was shown to *p*-dope semimetallic graphene[28]. This suggests that chloroform could serve as an effective *p*-type dopant for 2D semiconductors such as $WSe_2$, offering a simple and scalable approach compared to existing doping techniques. However, solvent-based doping is often regarded as transient, and the impact of chloroform doping on the electrical performance of *p*-channel $WSe_2$ transistors has not yet been studied.

In this work, we demonstrate that chloroform can induce high-performance, stable, and high-yield *p*-doping in monolayer $WSe_2$ transistors (**Figure 1b**). Exposing monolayer $WSe_2$ transistors to chloroform increases $I_D$ by two orders of magnitude, with hole currents reaching up to 203 μA/μm at $V_{DS}$ = -1 V (**Figure 1c**). These devices also maintain large $I_{on}/I_{off}$ ratios (~$10^{10}$) and a low $R_C$ of 2.5 kΩ·μm (at room temperature) and 1.0 kΩ·μm (at 10 K). Compared to recent approaches such as contact engineering (e.g. Sb/Pt[8,31], Ru[32,33]), oxide-based doping ($WO_x$[14,34], $MoO_x$[8,35], $NO_x$[15,34]), and other halide-based dopants ($HAuCl_4$[17], $RuCl_3$[19]), chloroform doping achieves one of the highest reported values for *p*-type transistor current (**Figure 1a**). Additionally, we observe that chloroform-doped transistors remain stable over 8 months (retaining 81% of initial doped $I_{D,max}$) and survive annealing at 150°C. Density functional theory (DFT) reveals that chloroform binds strongly (> 260 meV, i.e., > $10k_BT$ at 296 K) to $WSe_2$ without introducing mid-gap states. Atomic-force microscopy (AFM), Auger electron spectroscopy (AES), and X-ray photoelectron spectroscopy (XPS) suggest that chloroform intercalates at the $WSe_2$ interface with the gate oxide, further stabilizing its doping. This straightforward approach enhances *p*-type performance in $WSe_2$ transistors and complements other contact and interface engineering techniques for advancing 2D semiconductor technologies.

Optical spectroscopy provides insights into the physical and chemical interactions between the $WSe_2$ and chloroform. Raman spectra of monolayer $WSe_2$ soaked overnight in chloroform shows no significant changes in the E'/$A_1$' peak intensity ratio (**Figure 1d**), which suggests that long-term chloroform exposure does not significantly increase the $WSe_2$ defectivity. We also do not observe an increase in LA(M) or 2LA(M) peak intensity associated with disruption of the $WSe_2$ lattice (**Supplementary Fig. S1**). The photoluminescence (PL) spectrum after chloroform doping (**Figure 1e**) shows lower intensity than for the undoped sample. This PL quenching is consistent with a chloroform-induced increase in the hole concentration, leading to more non-radiative recombination via positive trions[29]. Additionally, the negligible change in surface roughness and morphology after doping indicates that residue adsorption does not play a significant role (**Supplementary Fig. S2**).

To investigate the electrical performance of chloroform-doped $WSe_2$, continuous monolayer CVD-grown $WSe_2$ was transferred onto an array of prefabricated ~5 nm $HfO_2$ local back gates (**Figure 2a**). The local back gates were defined by photolithography and lift-off of 2/8 nm Ti/Pt, followed by thermal atomic layer deposition of $HfO_2$ gate dielectric with equivalent oxide thickness of 1.23 nm. The $WSe_2$ channel was patterned by $XeF_2$ etching. Fine contact regions were also defined using e-beam lithography with a bilayer PMMA resist stack. Pd/Au (20/20 nm) were deposited by evaporation at ~ $10^{-7}$ Torr, followed by lift-off in acetone overnight, then rinsed in isopropanol (IPA). Electrical measurements were conducted in vacuum at ~$10^{-4}$ Torr. After initial device measurements, the devices were soaked in chloroform for > 8 hours and re-measured in vacuum (see Methods for more details).

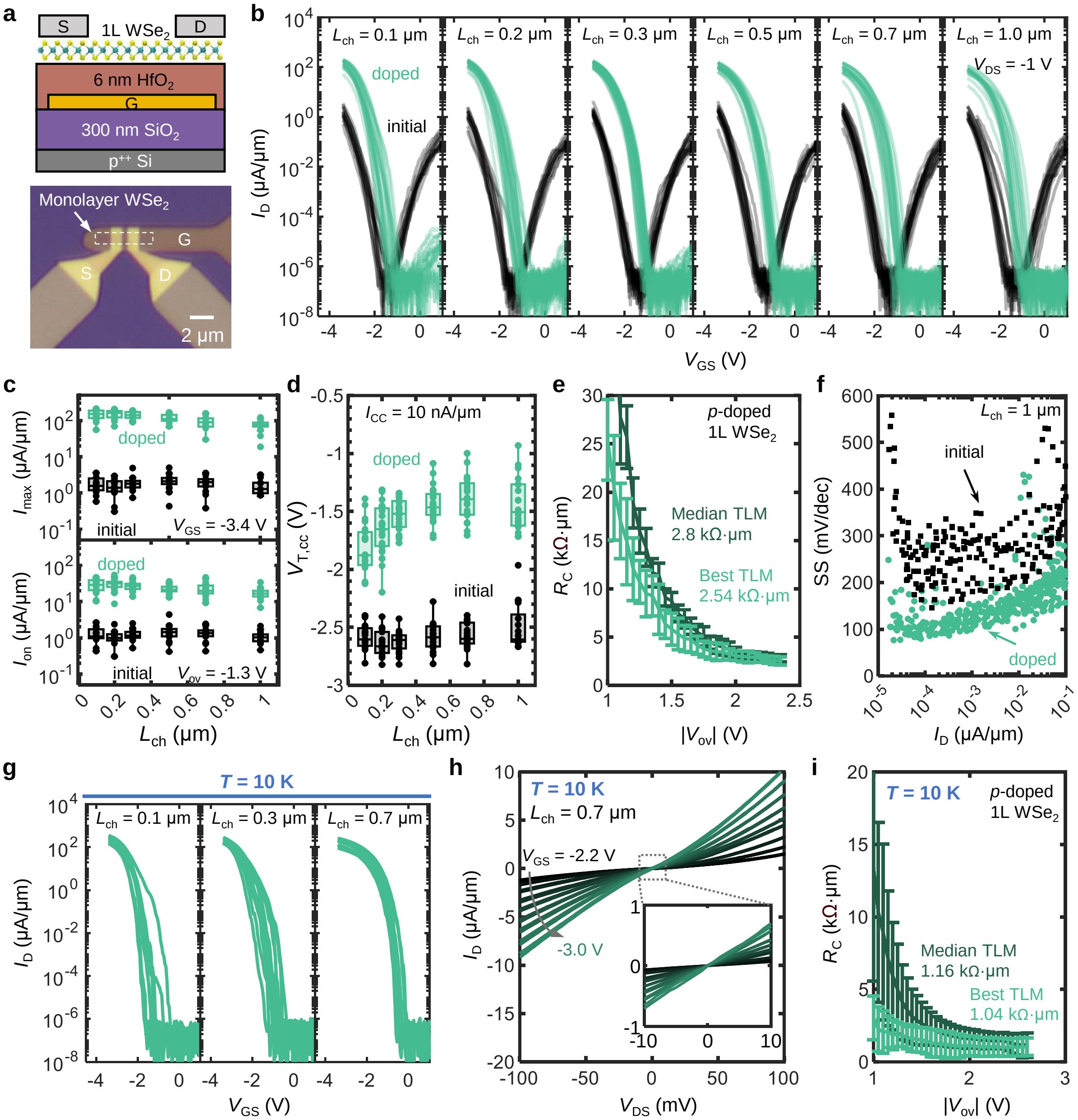

**Figure 2**. **Electrical Characterization of Chloroform-doped Monolayer $WSe_2$ transistors**. **a**, Cross-sectional schematic of $WSe_2$ transistor (top) and optical microscope image of fabricated device (bottom). **b**, Measured $I_D$ vs. $V_{GS}$ before and after doping at several channel lengths ($L_{ch}$) from 0.1 to 1 μm. **c**, $L_{ch}$-dependent statistical analysis before and after doping of (top) maximum drain-current $I_{D,max}$ at $V_{GS}$ = - 3.4 V and (bottom) on-state current $I_{on}$ at an overdrive voltage $V_{ov}$ = - 1.3 V. **d**, Threshold voltage ($V_{T,cc}$) at a constant current of 10 nA/μm before and after doping. All devices display a positive shift in $V_T$, indicating *p*-doping. **e**, Contact resistance ($R_C$) of chloroform-doped $WSe_2$ devices with Pd contacts, extracted using the transfer length method (TLM). **f**, Subthreshold swing (SS) vs. $I_D$ in $L_{ch}$ = 1 μm devices before and after doping. Doped devices show lower SS across the whole sub-threshold $I_D$ range. **g**, Measured $I_D$ vs. $V_{GS}$ at 10 K after doping for various channel lengths ($L_{ch}$ = 0.1 to 0.7 μm). **h**, $I_D$ vs. $V_{DS}$ curve for a representative $L_{ch}$ = 0.7 μm device at 10 K from $V_{GS}$ = - 3.0 V to - 2.2 V in steps of 0.1 V increments. The inset shows a magnified view of the low-voltage region. **i**, $R_C$ of chloroform-doped $WSe_2$ devices with Pd contacts at 10 K, extracted using the TLM method. Notably, a low contact resistance and high drain current is still maintained at cryogenic temperatures.

**Figure 2b** presents the transfer characteristics of 101 transistors before and after *p*-doping at $V_{DS}$ = -1 V, with channel lengths ($L_{ch}$) ranging from 100 nm to 1 μm. Prior to doping, the devices exhibit a highly negative $V_T$ around -2.6 V. After doping, the $V_T$ shifts positively and the maximum $I_D$ uniformly increases by ~100× across all devices, from ~ 1 μA/μm to >100 μA/μm (**Figure 2c**). The low device-to-device variation after doping demonstrates the reproducibility of this doping method.

We can further understand the origin of these improvements by examining the effects of chloroform upon the statistical distributions of the maximum hole current $I_{D,max}$ (at $V_{GS}$ = -3.4 V) and the on-state hole current $I_{on}$ at a fixed overdrive voltage ($V_{ov} = |V_{GS} - V_T|$), both shown in **Figure 2c**. Interestingly, devices of all channel lengths show concurrent increases in max $I_D$, and in $I_{on}$ at fixed $V_{ov}$ = 1.3 V. Evidently, the chloroform doping shifts $V_T$ positively, but the observed increase in $I_{on}$ at fixed $V_{ov}$ in both long- and short-channel devices suggests the improvement is a combined effect of increased mobility and reduced $R_C$. The maximum transconductance ($g_m$) of each device shows a 30.5× median increase (from 1.07 to 33 μS/μm) for $L_{ch}$ = 1 μm devices (**Supplementary Fig. S3a**), consistent with this interpretation.

All devices demonstrate a positive $V_T$ shift ($V_T$ extracted at a constant current 10 nA/μm)[36], with a median shift value of 1.0 V (from -2.6 V to -1.6 V), consistent with *p*-doping (**Figure 2d**). Additionally, unlike other SCTD methods[12,15,17], the chloroform-doped transistors did not exhibit any degradation in off-state current even at the shortest $L_{ch}$ of 100 nm. We extract the $R_C$ of our Pd-contacted doped monolayer $WSe_2$ devices using the transfer length method (TLM), yielding an $R_C$ of 2.5 (2.8) kΩ·μm for our best (median) pseudo-TLM structure (as described in Methods) (**Figure 2e**). In comparison, the initial $R_C$ before chloroform exposure was 168 kΩ·μm (**Supplementary Fig. S3b**). The improved $R_C$ likely stems from the doping of the $WSe_2$ region near the contacts, which narrows the metal-semiconductor energy barrier width and enhances the contribution of tunneling[37,38]. Notably, this $R_C$ value represents the best reported for Pd contacts on monolayer $WSe_2$ and is comparable to the highest-performing contact schemes reported to date (e.g. Sb/Pt with $MoO_x$ doping[31], $WO_x$ and NO doping[34]). **Table S1** benchmarks the performance of *p*-type monolayer $WSe_2$, highlighting that our devices achieve state-of-the-art $R_C$ and performance metrics.

Comparison of the subthreshold swing (SS) before and after doping reveals a decrease in SS for the doped devices, down to 81.4 mV/dec from 144 mV/dec at room temperature (**Figure 2f, Supplementary Fig. S3c**). This improvement in SS spans the entire subthreshold range of $I_D$ from doping. (**Figure 2f, Supplementary Fig. S3d**), which may be due to passivation of interfacial defects and could also explain the increase in mobility[39].

High $R_C$ also limits the operation of $WSe_2$ transistors at cryogenic temperatures, impeding the study of $WSe_2$ in quantum transport devices. Chloroform-doped monolayer $WSe_2$ transistors at 10 K (**Figure 2g**) show consistently high hole current across all devices, up to 403 μA/μm at $V_{DS}$ = -1 V for $L_{ch}$ = 0.1 μm (**Supplementary Fig. S4**), with relatively linear $I_D$ vs. $V_{DS}$ (**Figure 2h**). The cryogenic $R_C$ ~ 1.0 kΩ·μm was extracted from a pseudo-TLM fit to devices ranging from $L_{ch}$ = 100 nm to 1 μm (**Figure 2i**). To our knowledge, this is the lowest *p*-type $R_C$ reported to date for cryogenic temperatures.

Temperature-dependent PL measurements of $WSe_2$ before (**Figure 3a**) and after (**Figure 3b**) chloroform exposure provide additional evidence for charge transfer. In both samples, the peaks at 1.75 eV and 1.71 eV correspond to neutral excitons (X) and trions (T), respectively, with the 40 meV difference matching the reported trion binding energy[40,41]. Additionally, the relative intensity of the trion peak is greater than the exciton peak in the chloroform-exposed $WSe_2$ (**Supplementary Fig. S5**), as expected for increased doping[42].

In both doped and undoped $WSe_2$, the three lower-energy peaks -- labeled L1 (~1.67 eV), L2 (~1.64 eV), and L3 (~1.60 eV) -- resemble previous reports of excitonic bound states[42] which display sublinear excitation power dependence (**Supplementary Fig. S5**) and rapidly quench above 100 K[43]. These characteristics are consistent with weakly-bound defect or donor states near the valence band maxima[43,44]. The intensity of the L1, L2, and L3 peaks is significantly higher in the doped sample (**Figure 3a**,**b**), which indicates increased radiative recombination of electrons and holes bound to different sites and could be explained by the Fermi-level moving towards the valence band after chloroform doping[45,46].

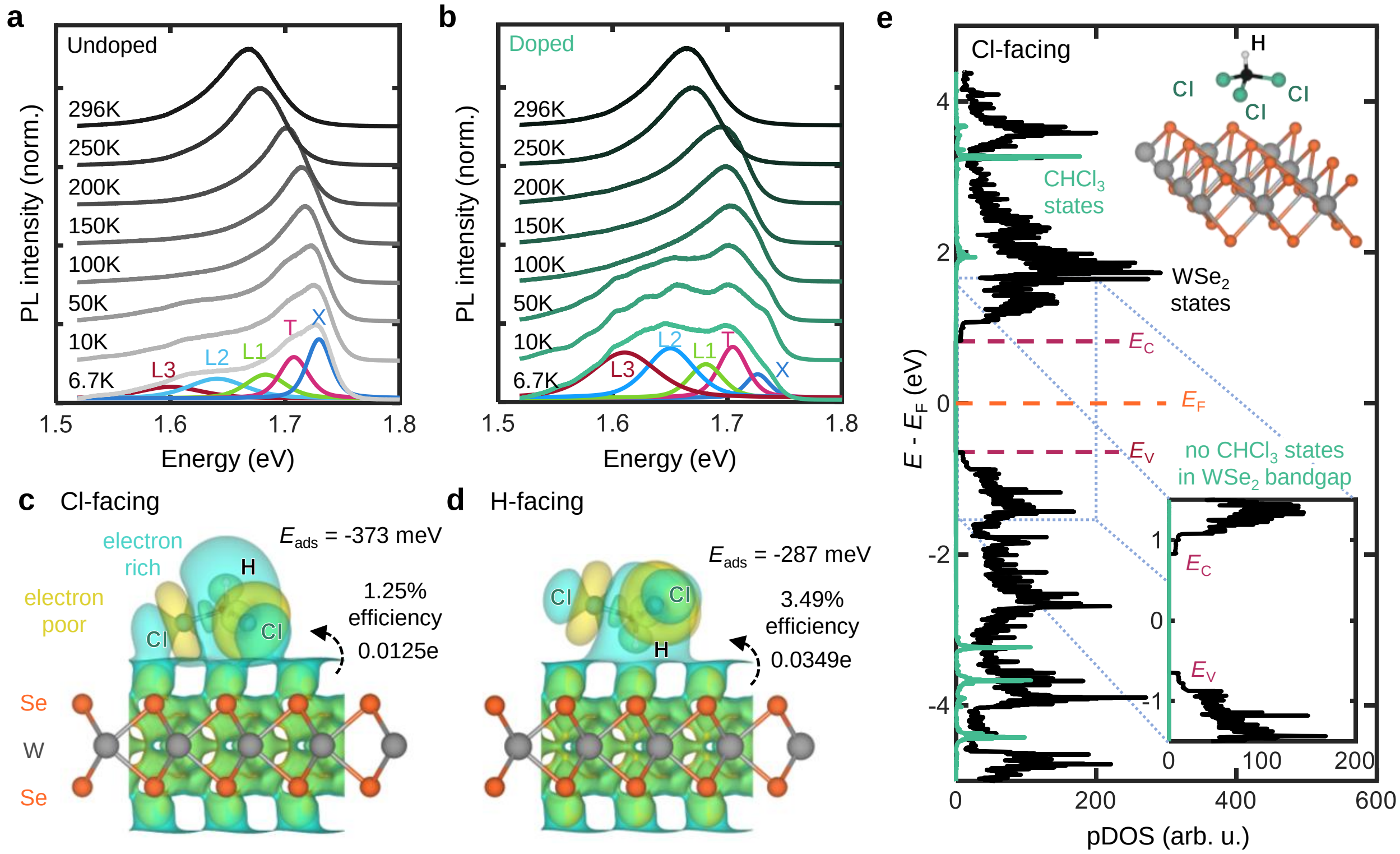

**Figure 3**. **Charger transfer mechanism of monolayer $WSe_2$ doped with chloroform. a**, Photoluminescence (PL) spectra of an undoped monolayer $WSe_2$ sample at different temperatures (6.7 to 296 K). **b**, PL spectra of chloroform-doped monolayer $WSe_2$ sample from 6.7 to 296 K. Representative Gaussian-Lorentzian blend curve fits are shown for the 6.7 K spectra in panel a and b, corresponding to the neutral exciton (X), trion (T), and L1-L3 peaks (described further in the text). Density functional theory (DFT) simulated isosurfaces of monolayer $WSe_2$ with adsorbed chloroform in **c**, Cl-facing and **d**, H-facing configurations, as well as the calculated adsorption energy ($E_{ads}$). The value of the Bader charge transfer efficiency is shown for each chloroform orientation, corresponding to the charge transfer of one chloroform atom. **e**, Projected density of states (pDOS) contributions from monolayer $WSe_2$ and chloroform to the overall DOS in the Cl-facing orientation. The Fermi energy $E_F$ is referenced to zero and marked with a dashed orange line; the valence band maximum $E_V$ and conduction band minimum $E_C$ are marked with dashed pink lines. Noticeably, no chloroform states are formed in the $WSe_2$ band gap. The inset shows a zoomed-in view of the PDOS contributions around the $WSe_2$ band gap.

To investigate the $p$-doping mechanism, we modeled the interactions between chloroform and $WSe_2$ using density functional theory (DFT). The chloroform absorption site was determined by relaxation against a rigid 5×5 $WSe_2$ supercell, considering geometries where the hydrogen atom faced towards (H-facing) or away from (Cl-facing) the monolayer $WSe_2$. Additional computational details are provided in the Methods section.

The Cl-facing and H-facing chloroform orientations exhibit favorable adsorption energies ($E_{ads}$) of -373 and -287 meV, respectively, indicating strong physisorption ($|E_{ads}| \gg k_B T$) to the $WSe_2$ without inducing covalent chemical modification (**Figure 3c**,**d**). This greatly exceeds adsorption energies between some small molecules and graphene ($|E_{ads}| < 100$ meV)[47] and are on the high end of values calculated for other adsorbates on TMDs (50 to 333 meV)[48,49].

Bader charge analysis reveals that the chloroform withdraws electrons from the $WSe_2$ in both configurations (**Figure 3c**,**d**). In both the Cl-facing and H-facing orientations, the adsorbed chloroform molecule gains a net charge of 0.0125 and 0.0349 excess electrons, respectively, confirming $p$-doping. These electron transfers are comparable to that between several well-known TMD SCTD systems, including: (i) $MoO_3$, a well-established $p$-dopant, and $MoS_2$ (~0.077 electrons[50] transferred from $MoS_2$ to $MoO_3$ per unit cell of $MoS_2$, assuming full surface coverage), (ii) $MoS_2$ and acetone (~0.039 electrons transferred to $MoS_2$ per molecule of acetone[51]), which is known to strongly $n$-dope 2D TMDs[29,30], and (iii) nitric oxide (NO) and $WS_2$ (0.018 electrons per molecule of NO[48], which is known to be an excellent $p$-dopant for 2D TMDs[15,34]. The calculated Bader charge transfer could be increased by up to a factor of four due to substrate interactions[51], which would further enhance the predicted efficiency of hole doping due to chloroform adsorption.

According to the atom-resolved projected density of states (pDOS), the chloroform orbitals are located more than 1 eV below the valence band edge or above the conduction band edge of monolayer $WSe_2$, and the chloroform molecule does not introduce electronic states near the band extrema or in the band gap (**Figure 3e**, **Supplementary Fig. S6**). This suggests that the charge transfer between chloroform and $WSe_2$ occurs without covalent bond formation or orbital hybridization, which is consistent with the adsorption energies we calculate for the chloroform/$WSe_2$ system (additional states associated with chemisorption are typically accompanied by an adsorption energy < -500 meV[52]). The absence of states formed in or near the band gap suggests that chloroform doping avoids introducing scattering sites that could degrade the mobility of $WSe_2$.

To evaluate the long-term stability of chloroform doping, we regularly measured the charge transport characteristics of doped monolayer $WSe_2$ transistors for more than 8 months. **Figure 4a** shows the evolution of $I_D$ vs. $V_{GS}$ sweeps for a single doped monolayer $WSe_2$ device over 243 days (> 8 months). The maximum drain current $I_{D,max}$ at $V_{GS}$ = -3.4 V slightly decreases after 243 days from 113 µA/µm to 97.4 µA/µm. **Figure 4b** plots the forward and backward sweep $V_{T,cc}$ for devices with $L_{ch}$ = 1 µm, revealing a median negative shift of -0.18 V over the course of long-term testing. This small negative shift in $V_T$ could indicate a slight reduction in $p$-doping due to chloroform desorption.

**Figure 4c** summarizes the evolution of $I_D$ in 1 µm long devices as $I_{D,max}$ for $V_{GS}$ = -3.4 V and at $V_{ov}$ = 1.5 V. Chloroform doping remains remarkably stable over time, with the median $I_{D,max}$ retaining > 96% after 6 days and > 81% after 243 days. We note that the decrease in $I_{D,max}$ can be partially attributed to the negative $V_T$ shift because the $I_{on}$ remains relatively stable. After 8 months, the final $I_{D,max}$ is still 76.1× higher than the initial undoped $I_{D,max}$ for the same set of devices. This demonstrates that the improved *p*-type performance from chloroform doping is highly stable over time. In contrast, other doping techniques (e.g. $MoO_3$[12,53], $O_3$ oxidation[54]) degrade rapidly in air, losing functionality over the course of several hours or days**.** The next-best reported example, nitric oxide, maintained performance after 24 days[55]. Additionally, the low $R_C$ from chloroform doping was maintained after 8 months (**Supplementary Fig. S7**). This stability of chloroform doping over time is consistent with the strong physisorption predicted from our DFT simulations (**Figure 3c,d**).

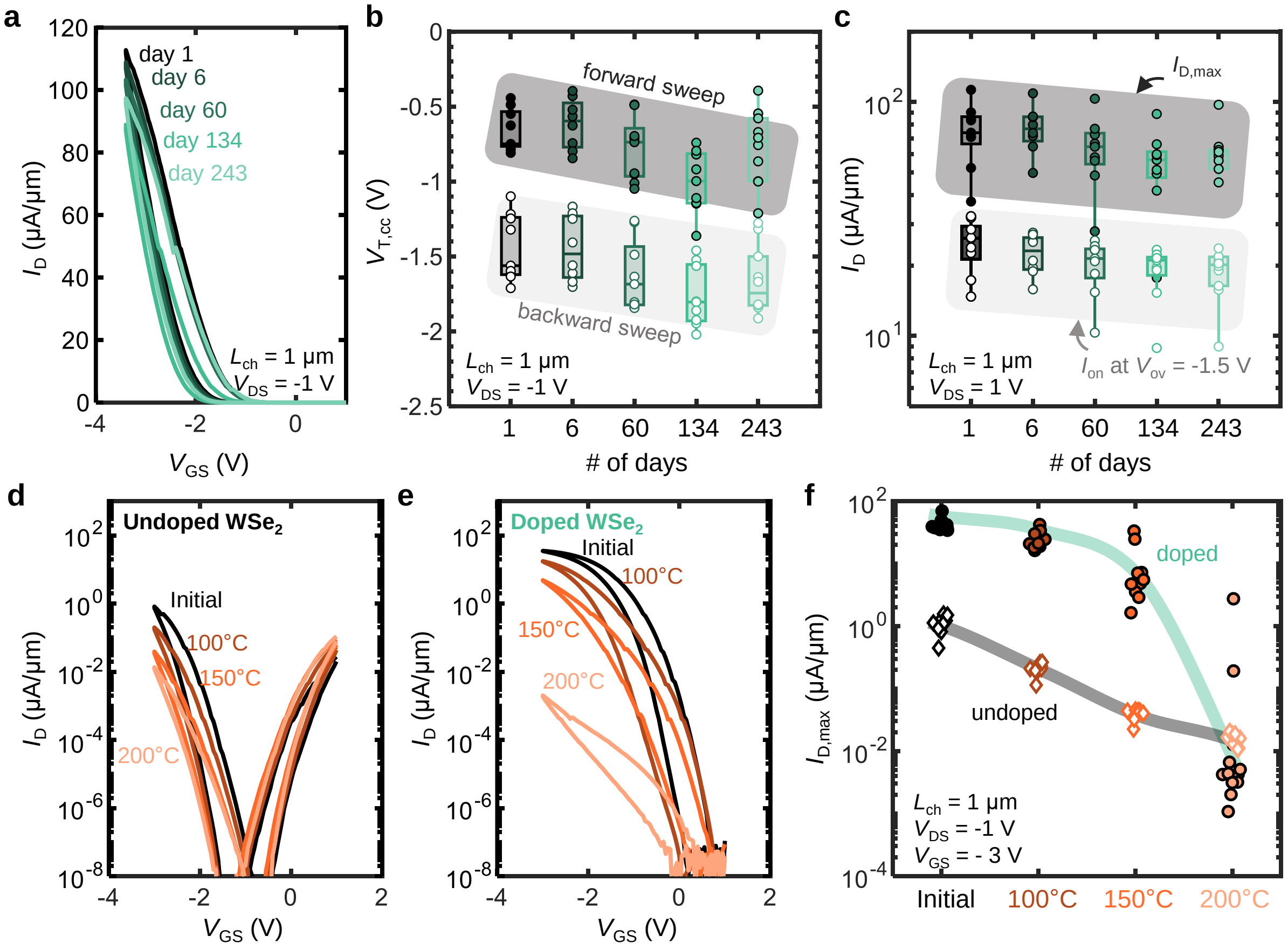

**Figure 4. Time (a-c) and Temperature (d-f) Stability of chloroform-doped $WSe_2$ devices. a,** $I_D$ vs. $V_{GS}$ curves of a $L_{ch}$ = 1 µm device immediately after doping and after 6, 60, 134 and 243 days. **b**, Threshold voltage ($V_{T,cc}$) vs. days after doping for all $L_{ch}$ = 1 µm devices. $V_{T,cc}$ is extracted at a constant-current of 10 nA/µm at $V_{DS}$ = -1 V for both forward and backward sweeps. **c**, Drain current (both $I_{D,max}$ at $V_{GS}$ = - 3.4 V and $I_{on}$ at $V_{ov}$ = - 1.5 V) vs. days post-doping for all $L_{ch}$ = 1 µm devices. **d**, $I_D$ vs. $V_{GS}$ curves of an undoped $WSe_2$ device ($L_{ch}$ = 1 µm) at $V_{DS}$ = -1 V initially, then after annealing at 100, 150 and 200°C. **e**, $I_D$ vs. $V_{GS}$ curves of a doped $WSe_2$ device ($L_{ch}$ = 1 µm) at $V_{DS}$ = -1 V initially, then after annealing at 100, 150 and 200°C. **f**, $I_{D,max}$ at $V_{GS}$ = - 3.0 V after various annealing temperatures, for undoped and doped devices. For the annealing process, the devices are sequentially annealed in vacuum at ~$10^{-4}$ Torr for 30 minutes at the given temperature. After annealing, the devices are cooled to room temperature for electrical measurement, then re-annealed at the next temperature.

Thermal stability is also critical to enable further processing. We examined the thermal stability of undoped and chloroform-doped $WSe_2$ transistors by sequentially annealing them in vacuum for 30 minutes at 100, 150°C, and 200°C. **Figure 4d** shows the $I_D$ vs. $V_{GS}$ of a control device, displaying a progressive decrease in hole current after each annealing step. This decline may result from the desorption of weakly-bound water molecules, which also contribute to *p*-doping[56,57]. **Figure 4e** plots the $I_D$ vs. $V_{GS}$ evolutions for a chloroform-doped device after the same annealing sequence. Similar to the undoped control device, the doped device exhibited a slight reduction in hole current after annealing at 100°C and 150°C. However, after the 200°C anneal, the $I_{D,max}$ of the doped device dropped significantly to $3\times10^{-3}$ μA/μm, comparable to the control device under similar annealing conditions. This suggests that chloroform desorbs at elevated temperatures, reverting the device to an undoped state.

A summary of $I_{D,max}$ across all annealing stages (i.e., the initial state and anneals at 100°C, 150°C, and 200°C) shows that the doped devices remain > 140× higher in $I_{D,max}$ after annealing at 100°C and 150°C, (**Figure 4f**, **Supplementary Fig. S8**). However, following the 200°C anneal, there was a sharp drop in hole current of the doped devices, consistent with the desorption of chloroform and a reversal to the undoped state. This sequential annealing procedure suggests that 150°C can be treated as a safe upper-bound for the thermal stability of chloroform doping on $WSe_2$, although faster thermal ramping and cooling may reveal a thermal budget for higher temperatures. In any case, this 150ºC thermal budget for chloroform stability enables compatibility with oxide encapsulation by atomic-layer deposition (ALD), which often occurs between 100 to 200°C. This may further enhance the stability of the *p*-type doping, enable fabrication of top-gated devices, and allow for concurrent application of other doping techniques, such as solid charge transfer layers (e.g., $MoO_x$, $WO_x$).

To clarify the mechanism and stability of chloroform doping, we investigated its location relative to the $WSe_2$. **Figure 5a** presents a 20×20 μm$^2$ atomic force microscopy (AFM) topography image of an exfoliated $WSe_2$ flake in which the thickness increases from 2 to over 10 layers. The bilayer (2L) to four-layer (4L) region were measured in 2×2 μm$^2$ scans before and after doping (**Supplementary Fig. S9**), yielding the height distributions shown in **Figure 5b**. The peaks mark the height of the $SiO_2$, 2L, 3L, and 4L $WSe_2$ regions. There is no noticeable change in spacing between the $WSe_2$ layers, but the height difference between the $SiO_2$ and 2L $WSe_2$ increases by > 0.15 nm. This suggests that chloroform does not intercalate between the $WSe_2$ layers, but rather that chloroform either: (i) inserts at the $SiO_2$/ $WSe_2$ interface; or (ii) adsorbs on top of every $WSe_2$ layer. X-ray diffraction (XRD) reveals that the interplanar spacing remained constant at 0.645 nm after doping (**Supplementary Fig. S10**), supporting the conclusion that chloroform does not intercalate between $WSe_2$ layers.

We measured the dependence of chloroform adsorption on $WSe_2$ thickness using Auger electron spectroscopy (AES). **Figure 5c** shows a scanning electron microscopy (SEM) image of the exfoliated $WSe_2$ flake from **Figure 5a**, while **Figures 5d-f** display AES elemental maps of Se, Si, and Cl. **Figure 5g** plots the AES signal intensities along the line in **Figure 5c**. As a surface sensitive technique with an Auger electron escape depth of approximately 5 to 50 Å, AES confirms that the Se signal intensity scales with the $WSe_2$ thickness (**Figure 5d**). In contrast, the Cl signal is negligible outside the $WSe_2$ region, peaks within the 2L $WSe_2$ terrace, and diminishes significantly for thicker $WSe_2$ layers (**Figure 5f**,**g**).

AFM height mapping indicates uniform increases in height across the 2L, 3L, and 4L $WSe_2$ regions after chloroform exposure (**Figure 5b**), while AES mapping shows the highest Cl signal in the 2L region (**Figure 5f,g**). This suggests that chloroform intercalates at the $WSe_2$/oxide interface, as reported for graphene on $SiO_2$[28], with thicker $WSe_2$ regions attenuating the AES signals from Cl beneath the $WSe_2$. This interfacial chloroform may enhance the $WSe_2$ device performance by increasing the oxide/$WSe_2$ separation and reducing the influence of interfacial oxide *n*-doping[58] and trap states – which may contribute to the observed reduction in SS for doped $WSe_2$ devices (**Figure 2f**). Notably, the correlation between the Cl signal and $WSe_2$ regions suggests that $WSe_2$ is necessary for chloroform adsorption (**Figure 5f**). XPS is consistent with this observation, detecting a Cl peak only in substrate regions covered by monolayer $WSe_2$ (**Supplementary Fig. S11**). In contrast, there is no apparent Cl peak in the bare substrate regions of chloroform-soaked samples.

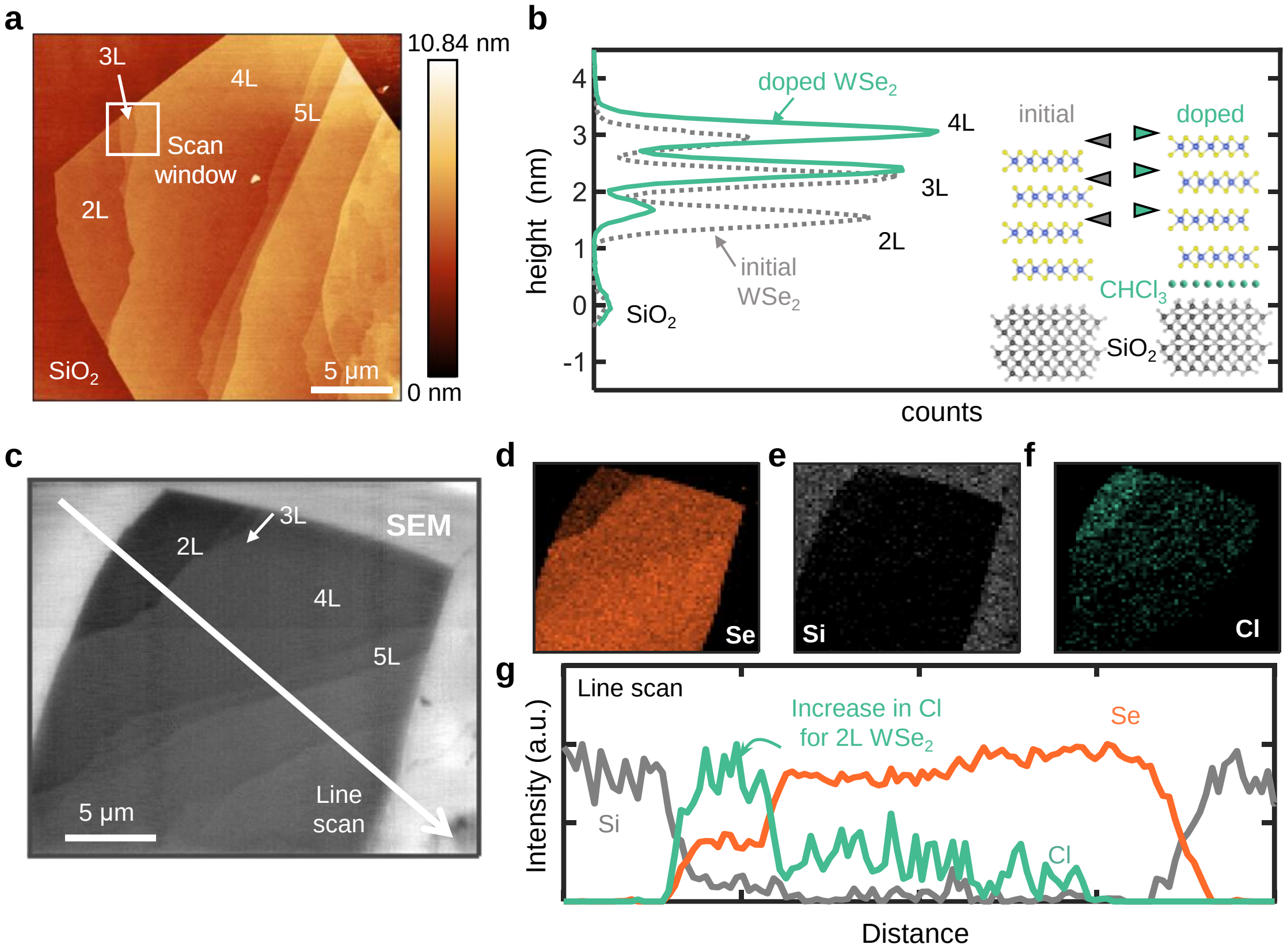

**Figure 5. Determination of chloroform location in a $WSe_2$/oxide stack. a,** Atomic force microscopy (AFM) of an exfoliated $WSe_2$ flake with various layer thicknesses. **b**, Height distribution of the exfoliated flake before (gray) and after (green) doping, in the 2-4L region (as marked in panel a). The peaks and triangles mark the height of the $SiO_2$, 2L, 3L, and 4L $WSe_2$ regions. There is no noticeable change in spacing between $WSe_2$ layers, but the difference between $SiO_2$ and 2L $WSe_2$ increases. **c**, Scanning electron microscope (SEM) image of the exfoliated $WSe_2$ flake as seen in panel a. **d-f**, Elemental mapping by Auger electron spectroscopy (AES) of a doped $WSe_2$ flake of Se, Si, and Cl respectively. The brighter pixels correspond to regions with higher elemental content. **g**, Line scan of elemental Se, Si, and Cl content extracted from panel d, showing an increase in Cl signal in the 2L $WSe_2$ region.

Overall, this work presents a straightforward and stable *p*-doping method to achieve high performance monolayer $WSe_2$ transistors, while providing new mechanistic insights into solvent-based doping techniques. By achieving significant improvements in hole current, $R_C$, and device stability, this method offers a viable path for future low-power 2D semiconductor applications.

## METHODS:

**Doping Process.** The $WSe_2$ sample was immersed in room temperature chloroform (SIGMA-Aldrich, No. 650498), in a watchglass-covered borosilicate beaker. Unless indicated otherwise, the doping process occurred overnight (> 8 hours). For device measurements, the doping process was performed after the initial device fabrication process was completed.

**Material Characterization.** Raman measurements were taken on the Horiba Labram HR Evolution Raman system in the Stanford Nanofabrication Shared Facility, using 532 nm laser excitation at 1% nominal laser power (120 μW) and a spot size < 1 μm in diameter. These parameters were selected to ensure minimal sample heating during measurement. For Raman and PL, solid-source chemical vapor deposition (CVD) monolayer $WSe_2$ grown on sapphire was transferred onto 100 nm $SiO_2$ before measurement. XPS was carried out using a PHI VersaProbe 4, equipped with a monochromatized Al Kα source (1486 eV) with a beam power of 50 W and beam energy of 15 kV, based pressure of $1.2 \times 10^{-7}$ Pa, and pass energy of 224 eV (step size: 0.8 eV) and 55 eV (step size: 0.1 eV) for survey and high-resolution acquisitions, respectively.

Bulk $WSe_2$ crystals were exfoliated with scotch tape onto oxygen-plasma cleaned silicon wafers with 100 nm thermal oxide. The exfoliated $WSe_2$ was probed for Auger electron spectroscopy (AES), X-ray diffraction (XRD), and the atomic force microscopy (AFM) images in **Supplementary Fig. S9**. Auger electron spectroscopy mapping, composition analysis and line scans on exfoliated $WSe_2$ were performed on a PHI 700 Scanning Auger Nanoprobe. XRD measurements were conducted using a PANalytic Empyrean system with a Cu-Kα source. Exfoliated $WSe_2$ flakes were probed with symmetric 2θ/ω scans. AFM was conducted on both the exfoliated $WSe_2$ and on CVD-grown $WSe_2$ on sapphire using a Bruker Dimension Icon in peak force mode with a NSC19 Al BS probe (nominal spring constant = 0.5 N/m).

**Local back-gate Device Fabrication on $HfO_2$ and Electrical Measurements.** Continuous 2-inch CVD-grown monolayer $WSe_2$ on sapphire was purchased from 2D semiconductors and transferred onto local back-gates of 5.3 nm $HfO_2$. The local back gates were defined by lift-off 2 nm/8 nm Ti/Pt followed by the $HfO_2$ gate dielectric by thermal atomic layer deposition at 200 °C. Coarse contact pads were then defined by lift-off 2/20 nm Ti/Pt. Polystyrene (PS) was spin-coated on top of the $WSe_2$ and then transferred in DI water. An $O_2$ plasma treatment (100 W, 1 min) of the $HfO_2$ dielectric was done before transferring the PS/$WSe_2$ film to modify the substrate's surface energy. The PS was then removed in toluene. Channel definition was done using electron beam lithography and etched by $XeF_2$ (2.5 T, 30 s, 3 cycles) to define a channel width of 1 μm. Electron beam lithography was used to pattern the fine contacts. Pd/Au (20/20 nm) was e-beam evaporated at $\sim 10^{-8}$ Torr. Electrical measurements were performed at 296 K in a Janis ST-100 vacuum probe station at $\sim 10^{-4}$ Torr, using a Keithley 4200 semiconductor parameter analyzer.

Cryogenic measurements were conducted in a Lakeshore cryoprobe station at ~ $10^{-6}$ Torr, using a Keithley 4200 semiconductor parameter analyzer. The sample was slowly cooled and left to stabilize overnight at 10 K before electrical testing.

For contact resistance ($R_C$) extraction, a pseudo-transfer length method (TLM) was used, as devices made were single devices with varying channel lengths. In this method, all devices at a certain channel length were used for $R_C$ extraction. The total resistance in kΩ·μm (normalized by the channel width) can be expressed as $R_{TOT} = 2R_C + R_{ch} = 2R_C + R_{sh}L_{ch}$, where $R_{sh}$ is the sheet resistance of the channel and $R_{ch}$ is the channel resistance. $R_C$ is evaluated by plotting $R_{TOT}$ versus $L_{ch}$ and drawing a linear fit through *all* data points, and the *y*-intercept at $L_{ch} = 0$ gives $2R_C$. The $R_C$ is extracted for each gate overdrive $V_{ov} = V_{GS} - V_T$, with $V_T$ from the constant-current method at $I_D = 10^{-2}$ μA/μm.

**Low Temperature Photoluminescence.** Low-temperature photoluminescence spectroscopy was conducted with a 532 nm excitation laser, ∼1 μm spot size, and 600 l/mm spectrometer grating. The laser power was fixed at 60 μW, unless otherwise noted. The emission was collected using a 50× objective with a numerical aperture of 0.55, with 2 second acquisition times and 2 accumulations. The sample was cooled to a base temperature of ~6.7 K, then was warmed up using a resistive heater for temperature dependent measurements. For temperature dependent experiments, the sample sat for 30 minutes at the desired temperature to stabilize before collecting the spectra. For this experiment, CVD-grown $WSe_2$ was wet transferred (as described above) onto 100 nm $SiO_2/p^{++}$ Si, then half of the chip was cleaved and subjected to an overnight chloroform soak. Several spots across both the control and doped sample were examined to ensure peak shape consistency. Finally peak fitting was conducted in Origin, using a Gaussian-Lorentzian blend.

**Vacuum Annealing Procedure and Electrical Testing.** Initial electrical measurements were performed at 296 K in a Janis ST-100 vacuum probe station at $\sim 10^{-4}$ Torr, using a Keithley 4200 semiconductor parameter analyzer. The samples were then in-situ annealed at 100°C, held for 30 minutes, then left to cool down. Measurements after annealing were taken once the devices cooled down (> 5 hours), at 300 K in vacuum. This process was then repeated at 150°C and 200° C respectively, with device measurements in between, without breaking vacuum.

**Density Functional Theory (DFT) Simulations.** First, a variable cell relaxation was performed to optimize the lattice coordinates within the monolayer $WSe_2$ primitive cell. The optimized primitive cell was then scaled to a 5 × 5 supercell, interfaced with a chloroform molecule, and then subjected to a fixed cell relaxation to determine the $WSe_2$/chloroform atomic coordinates. Both the Cl-facing and H-facing orientations were considered, where the chlorine or hydrogen atom of the chloroform molecule was oriented towards the $WSe_2$. Quantum ESPRESSO 7.1[59] was used for all DFT simulations, and the van der Waals interactions between the chloroform and $WSe_2$ monolayer were modeled using the DFT-D2 correction. The $WSe_2$ primitive cell was relaxed on a 25 × 25 × 1 *k*-point grid, and the relaxations and self-consistent calculations for the $WSe_2$/chloroform assemblies were performed on 5 × 5 × 1 *k*-point grids. We perform non-self-consistent calculations on a 15 × 15 × 1 *k*-point grid prior to extracting the density of states for the $WSe_2$ + chloroform assemblies. All DFT calculations use projector-augmented wave pseudopotentials with kinetic energy cutoffs and charge density cutoffs of 60 and 480 Ry, respectively. We use the “Bader” code[60] for Bader charge analysis, PyProcar[61] for plotting projected density of states and band structures, and VESTA[62] for plotting isosurfaces.

**Acknowledgements.** This work was supported by the Sandia Microelectronics: Accelerating Research Talent (SMART) Internship and TSMC under the Stanford SystemX Alliance. L.H., R.K.A.B., T.P., and E.P. also acknowledge the support of SUPREME, one of seven centers in JUMP 2.0, a Semiconductor Research Corporation (SRC) program sponsored by DARPA. R.K.A.B. acknowledges support from the Stanford Graduate Fellowship and NSERC PGS-D programs. Part of this work was performed at the Stanford Nanofabrication Facility (SNF) and Stanford Nano Shared Facilities (SNSF), supported by the National Science Foundation under award ECCS-2026822. The authors thank Krishna C. Saraswat, Marc Jaikissoon, and Alex Shearer for helpful discussions.

**Author Contributions**. L.H. fabricated the devices and conducted the device measurements and analysis under the supervision of A.J.M and E.P.. L.H. conducted the AFM, Raman, and Auger measurements. R.K.A.B. performed the DFT simulations. A.T.H. performed the CVD $WSe_2$ material growth. T.P. and L.H. performed the low temperature PL characterization and analysis with the help of A.P.S. under the supervision of F.L.. Z.Z. performed the XRD with L.H.. M.H. performed the $WSe_2$ exfoliation. L.H. and A.J.M. wrote the paper. All authors have given approval to the final version of the manuscript.

**Supplementary Information**

Figure S1 to S12

Table S1

Supplementary Information

# Low Resistance *P*-Type Contacts to Monolayer $WSe_2$ through Chlorinated Solvent Doping

Lauren Hoang[1], Robert K.A. Bennett[1], Anh Tuan Hoang[2], Tara Pena[1], Zhepeng Zhang[2], Marisa Hocking[2], Ashley P. Saunders[3], Fang Liu[3], Eric Pop[1,2,4], and Andrew J. Mannix[2,*]

[1]*Dept. of Electrical Engineering, Stanford University, Stanford, CA 94305, U.S.A.*

[2]*Dept. of Materials Science and Engineering, Stanford University, Stanford, CA 94305, U.S.A.*

[3]*Dept. of Chemistry, Stanford University, Stanford, CA 94305, U.S.A.*

[4]*Dept. of Applied Physics, Stanford University, Stanford, CA 94305, U.S.A.*

**This file includes**

Supplementary Figures S1-S12
Table S1

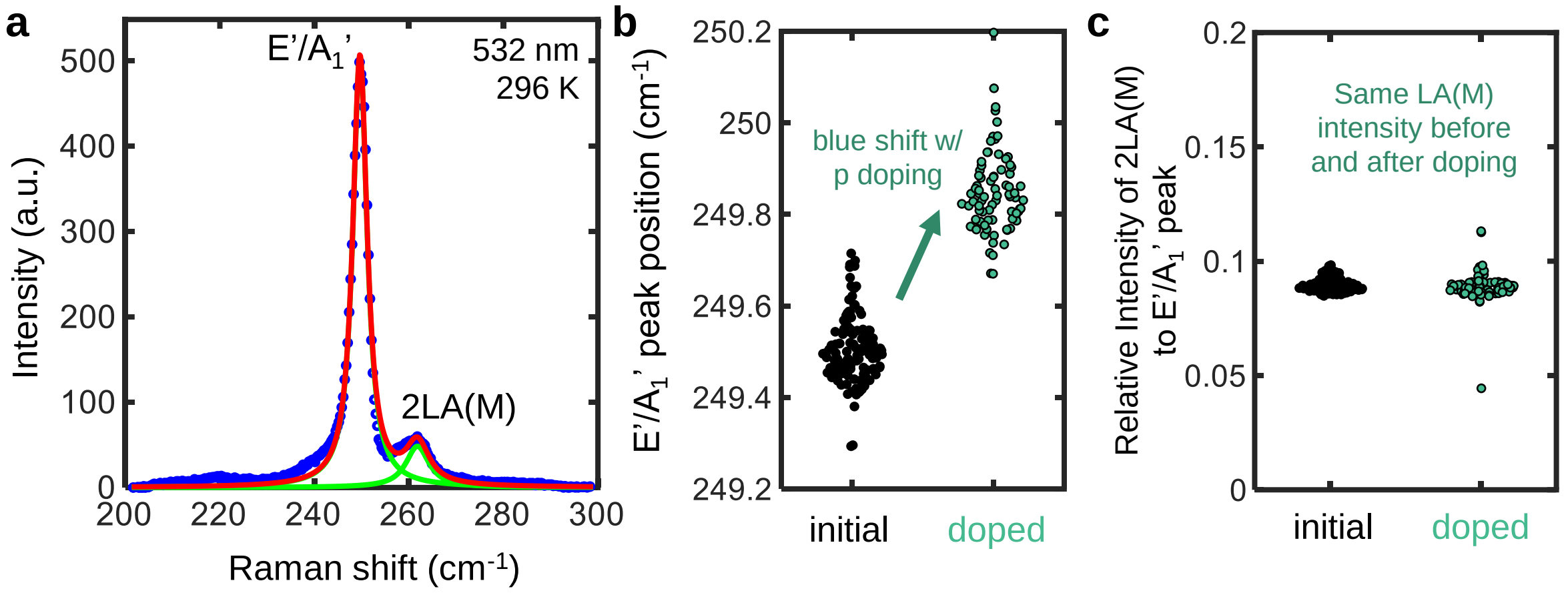

**Supplementary Fig. S1 | Raman Spectroscopy on Monolayer $WSe_2$. a**, Representative Raman spectra (blue points) and peak fitting of E'/$A_1$' and 2LA(M) peak using a Gaussian-Lorentzian line shape for all peaks. An iterative least-square method in MATLAB was used where the baseline of the spectra was subtracted prior to fitting. **b**, Extracted $WSe_2$ E'/$A_1$' peak position before and after chloroform doping, showing a blueshift[1–3] with doping. CVD-grown monolayer $WSe_2$ on $SiO_2$ was used for before and after doping comparison on the same $WSe_2$ flake. **c**, Relative intensity of the 2LA(M) peak to the E'/$A_1$' peak of monolayer $WSe_2$, showing negligible change in 2LA(M) intensity.

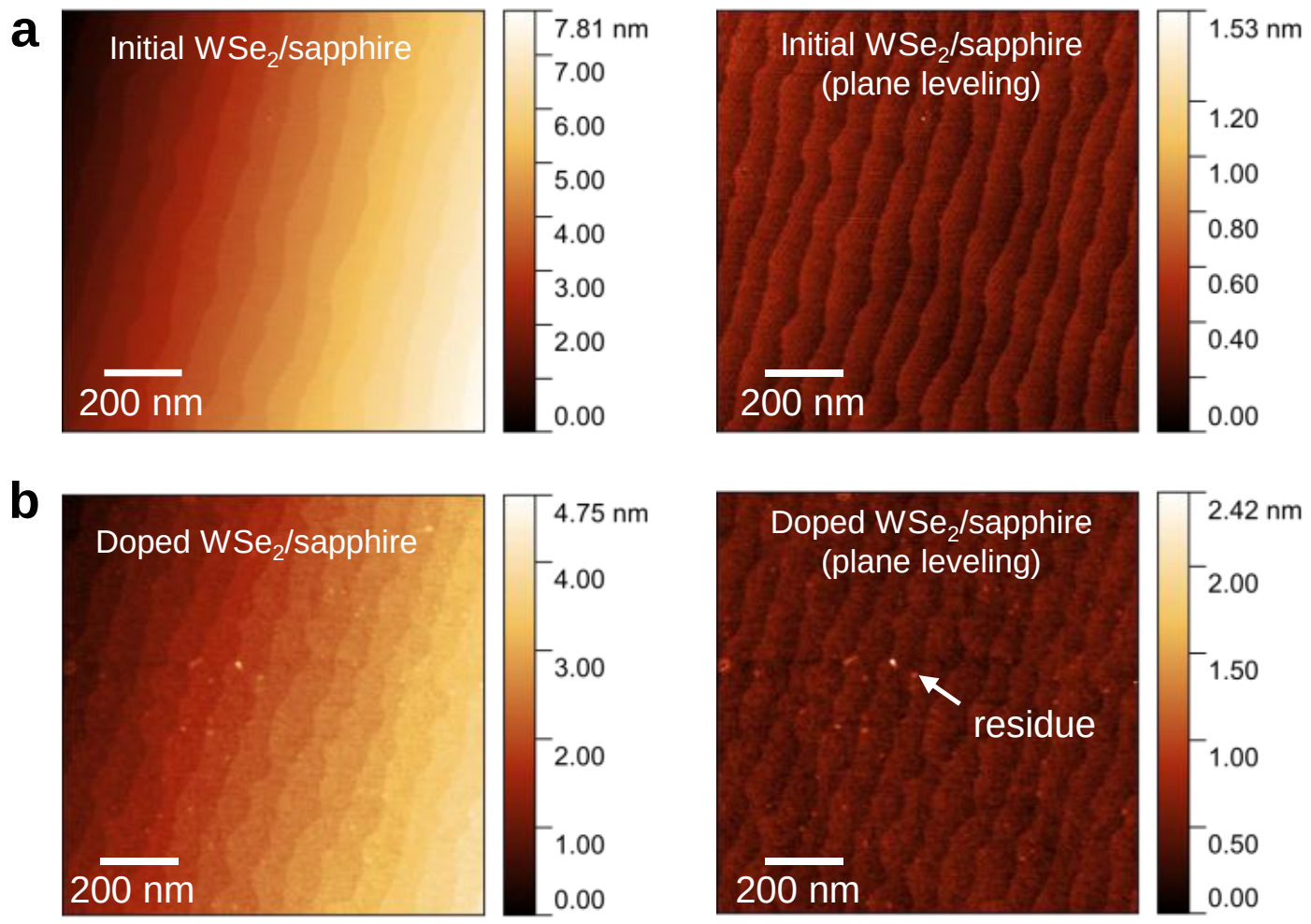

**Supplementary Fig. S2 | a**, Atomic Force Microscopy (AFM) image (scan size 1 × 1 μm$^2$) of monolayer $WSe_2$ grown directly on sapphire, showing low root mean square (RMS) surface roughness (RMS ~ 0.08 nm). **b**, AFM image (scan size 1 × 1 μm$^2$) of monolayer $WSe_2$ grown directly on sapphire, after doping in chloroform overnight. Some small residues can be identified on the $WSe_2$ flake, but there is still low surface roughness (RMS ~0.13 nm).

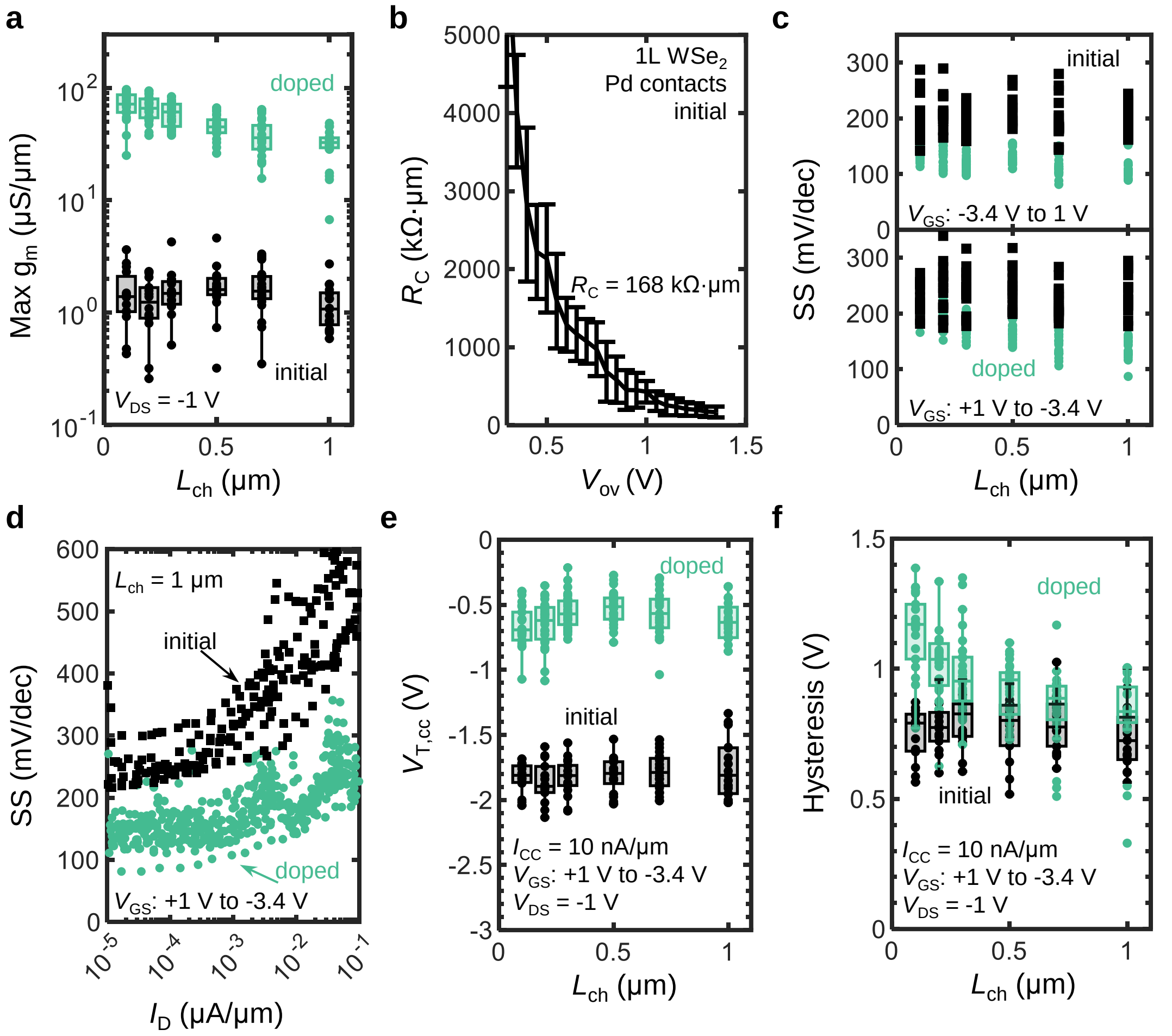

**Supplementary Fig. S3 | Additional Electrical Device Analysis. a**, Maximum transconductance ($g_m$) vs. channel length ($L_{ch}$) for undoped and doped devices, reaching >100 μS/μm for doped $WSe_2$. **b**, Contact resistance ($R_C$) vs. overdrive gate voltage ($V_{ov} = |V_{GS} - V_T|$) of monolayer $WSe_2$ devices with Pd contacts before doping. **c**, Minimum subthreshold swing (SS) with respect to $L_{ch}$ of doped and undoped monolayer $WSe_2$ devices for backward (top) and forward (bottom) gate voltage sweep directions. **d**, SS vs. $I_D$ in 1 μm long devices before and after doping, sweeping $V_{GS}$ from +1 V to -3.4V (forward direction). Doped devices show lower SS for the whole sub-threshold $I_D$ range. **e**, Constant-current threshold voltage ($V_{T,cc}$) at $I_{D,cc}$ = 10 nA/μm before and after *p*-doping, sweeping from +1 V to -3.4 V. **Figure 2d** shows the counterpart $V_{T,cc}$ for the backward sweep direction from -3.4 V to 1V. **f**, Hysteresis at $I_D$ = 10 nA/μm with respect to $L_{ch}$, for both undoped and doped devices.

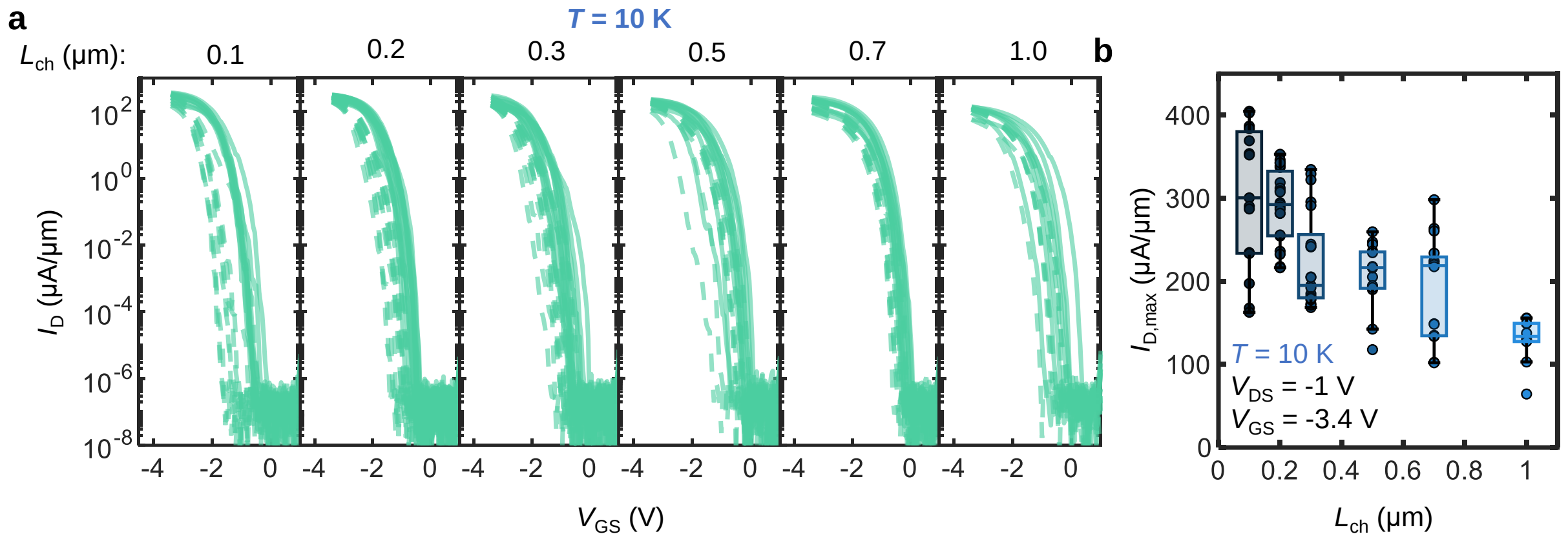

**Supplementary Fig. S4 | Low Temperature Electrical Measurements of Chloroform-doped Monolayer $WSe_2$ Transistors**. **a**, Measured $I_D$ vs. $V_{GS}$ at 10 K after doping for various channel length ($L_{ch}$ = 0.1 to 1.0 μm) devices. Solid (dashed) lines indicate gate voltage sweep from positive (negative) to negative (positive) gate bias. **b**, Maximum drain-current $I_{D,max}$ at $V_{GS}$ = - 3.4 V at 10 K as a function of $L_{ch}$. An $I_{D,max}$ of 403 μA/μm was achieved for a $L_{ch}$ = 0.1 μm device.

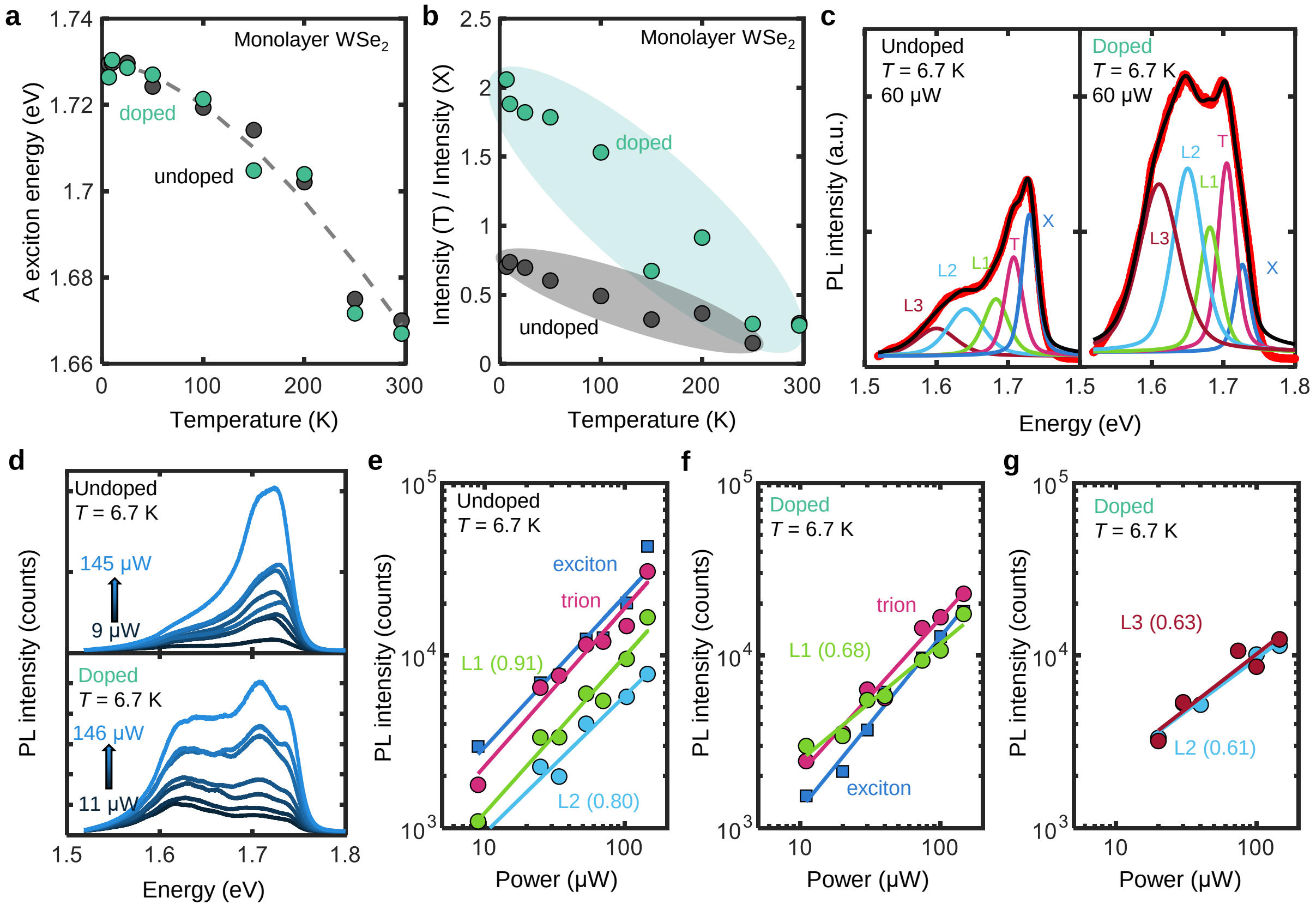

**Supplementary Fig. S5 | Temperature-dependent Photoluminescence Spectroscopy. a**, Neutral A exciton (X) energy as a function of temperature for both undoped and doped $WSe_2$. The exciton peak redshifts with increasing temperature, consistent with the expected band gap reduction described by the Varshni equation[4]. The data is fitted to the Varshni equation and displayed with a dashed line. **b**, Relative intensity of the trion peak to the exciton peak at different temperatures for undoped and doped $WSe_2$. The doped sample shows consistently higher trion intensity compared to the undoped sample. **c**, PL spectra and peak fitting of 5 peaks (X, T, L1, L2, L3) of an undoped (left) and doped (right) spectra. Noticeably, the exciton (X) intensity is higher than the trion (T), L1, L2, and L3 in the undoped sample, but lower in the doped sample. **d**, PL spectra under different laser powers collected at 6.7 K for an undoped (top) and doped (bottom) $WSe_2$ sample, respectively. **e**, PL intensity vs. laser power for the X, T, L1, and L2 peaks for undoped $WSe_2$. **f**, PL intensity vs. laser power for the X, T, and L1 peaks of doped $WSe_2$. The trion intensity is higher than the exciton intensity for doped $WSe_2$. **g**, PL intensity vs. laser power of L2 and L3 peaks of doped $WSe_2$. For panels e-g, the number in parenthesis indicates the power dependence (α) by fitting a power law $I \propto P^{\alpha}$ to the data, where $I$ is the PL peak intensity for a given laser power, $P$. The two higher energy peaks (X, T) for both samples have near-linear fit of the peak intensity with laser power ($\alpha \sim 1$), which is consistent with the exciton and trion behavior and is attributed to radiative recombination of excitons and trions. In comparison, peaks L1 - L3 exhibit sublinear power dependence, so we attribute their origin to bound excitons. We note that other peaks such as biexcitons and dark excitons have been reported to appear within this lower energy range[5,6]. The sublinear power dependence is indicative of radiative combination of electrons and holes separately localized at different spatial sites[7]. L1 - L3 peak intensities rapidly quench above 100 K as thermal stimulation perturbs the weak interaction between the defect bounded excitons[8].

**a**

Mulliken Population Analysis

| Structure | W | Se | $CHCl_3$ |
|---|---|---|---|
| Pristine $WSe_2$ | 12.970 | 16.515 | - |
| Cl-facing $WSe_2$ | 12.967 | 16.516 | 26.0125 |
| H-facing $WSe_2$ | 12.953 | 16.523 | 26.03486 |

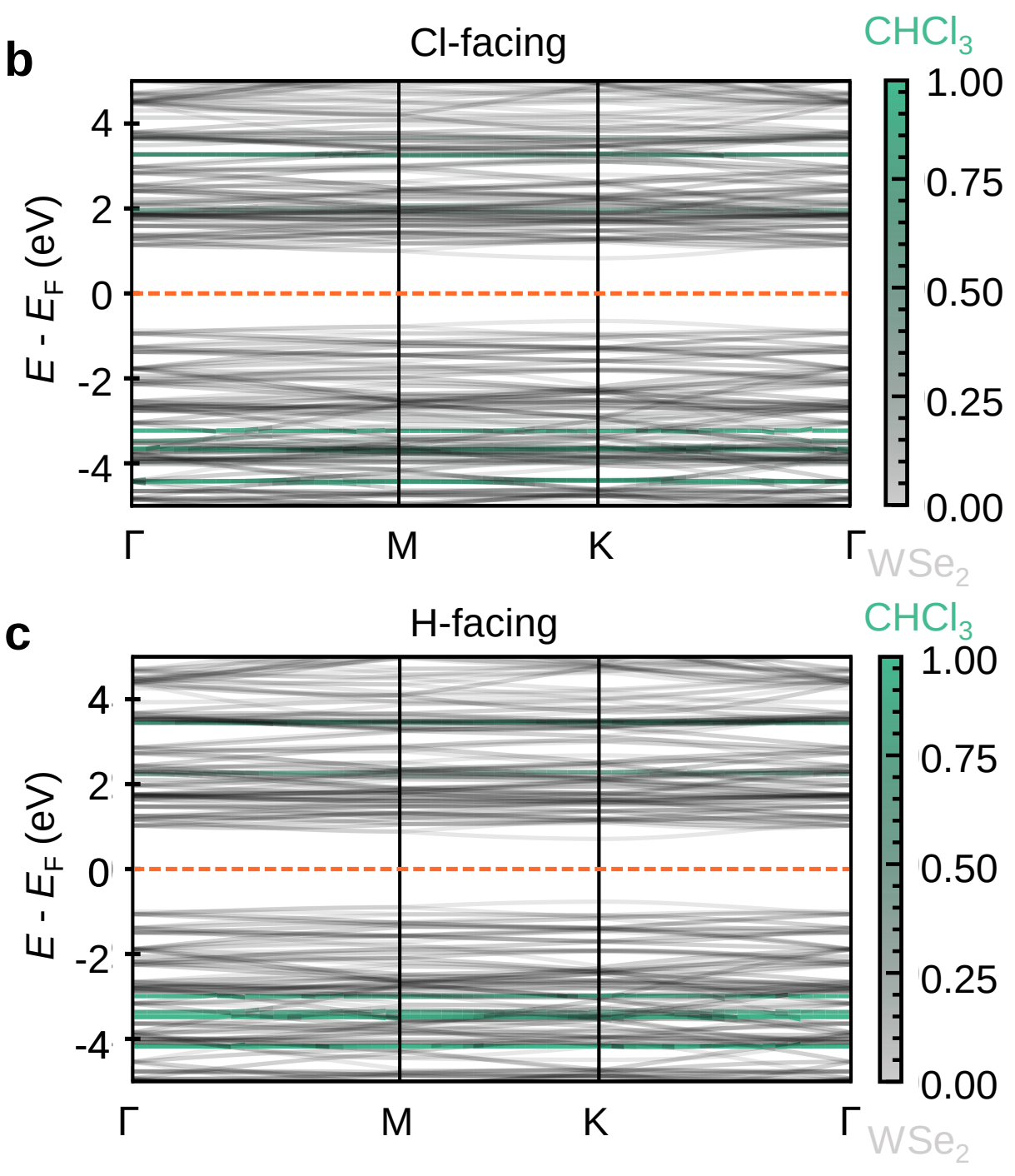

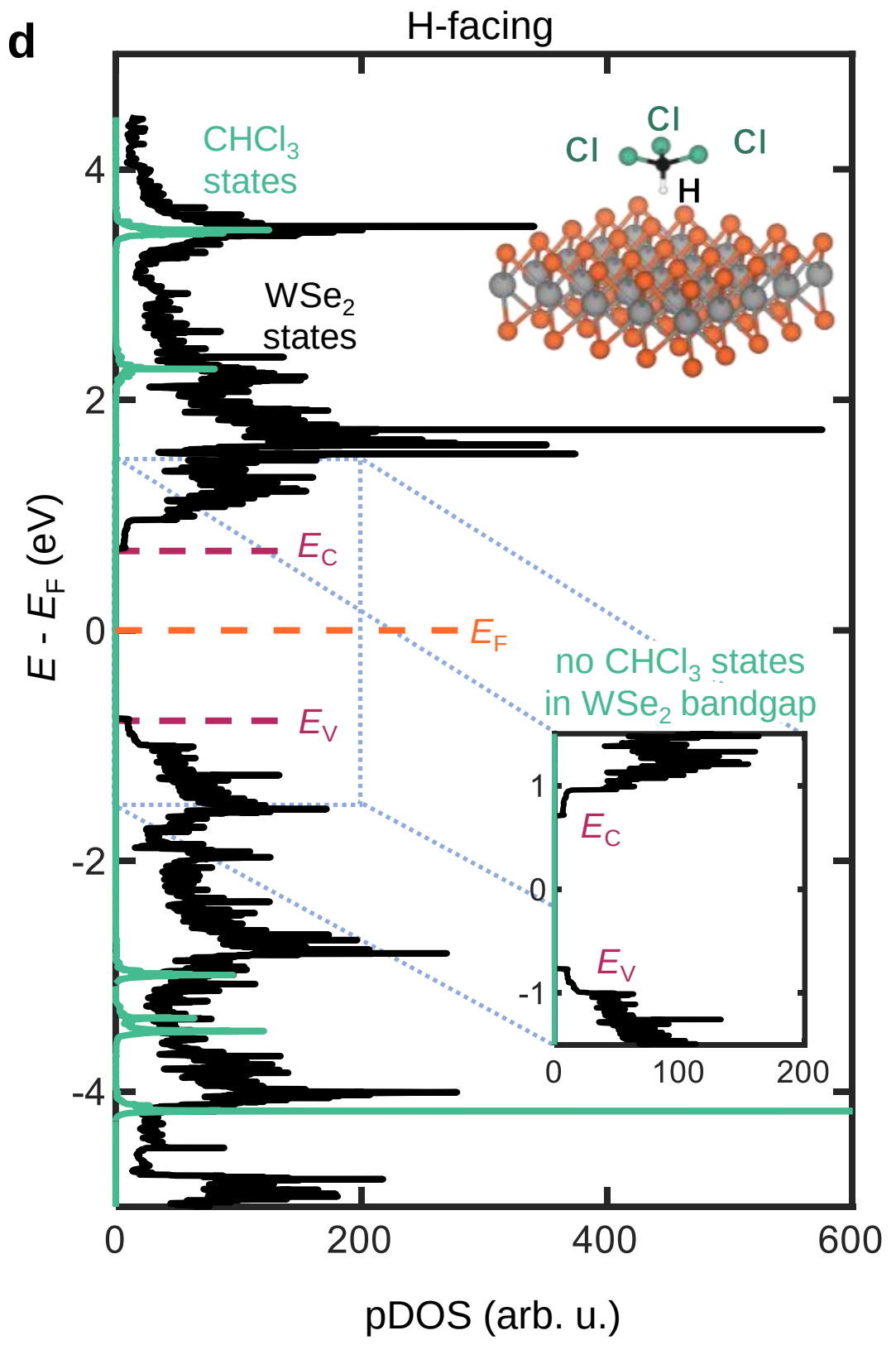

**Supplementary Fig. S6 | Density Functional Theory Simulations. a,** Mulliken population analysis of pristine and doped (Cl-facing and H-facing) monolayer $WSe_2$. The decrease in electron population of doped $WSe_2$ illustrates the *p*-doping of chloroform. Band structure of **b**, Cl-facing chloroform and **c**, H-facing chloroform interfaced with monolayer $WSe_2$. Transparent gray bands are contributions from monolayer $WSe_2$ and green bands are contributions from chloroform states. No chloroform states are formed in the $WSe_2$ band gap for both orientations. **d**, Projected density of states (pDOS) contributions from monolayer $WSe_2$ and chloroform to the overall DOS in the H-facing orientation. The Fermi energy $E_F$ is referenced to zero and marked with an orange line; the valence band maximum $E_V$ and conduction band minimum $E_C$ are marked with dashed pink lines. The inset shows a zoomed-in view of the PDOS contributions around the $WSe_2$ bandgap.

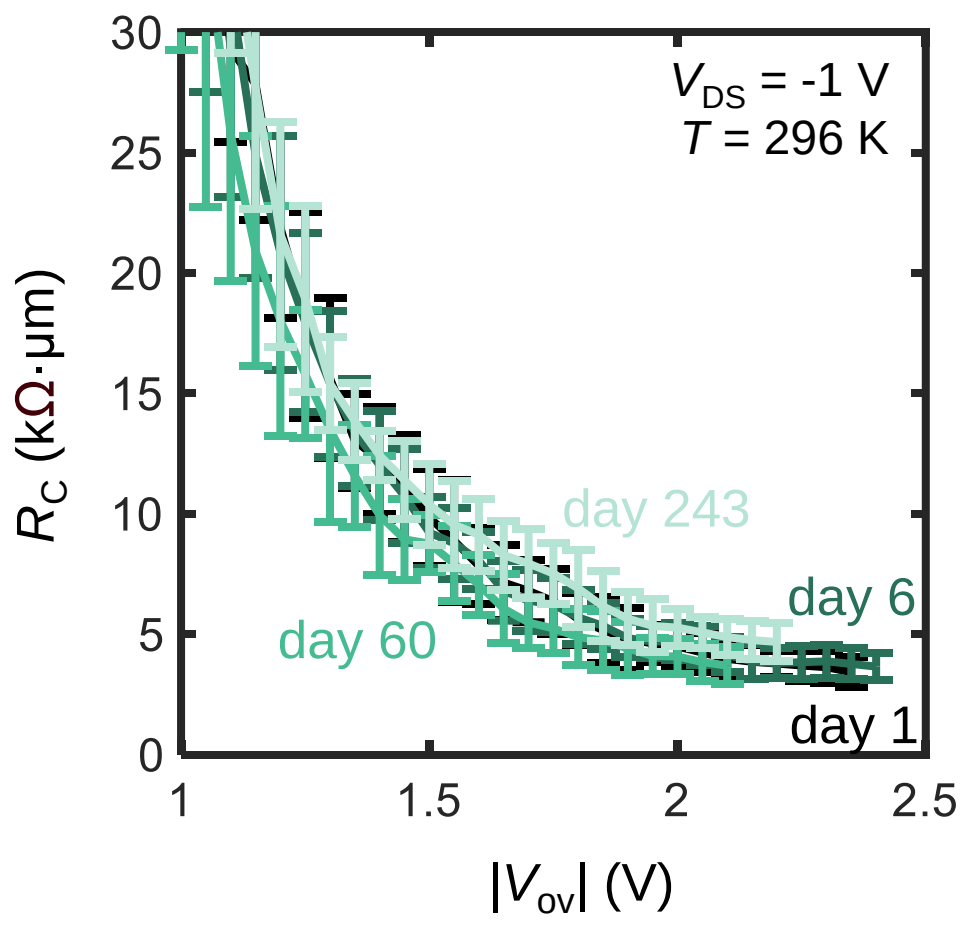

**Supplementary Fig. S7 |** Contact resistance ($R_C$) of chloroform-doped monolayer $WSe_2$ devices with Pd contacts 1 day, 6 days, 60 days, and 243 days (>8 months) after doping.

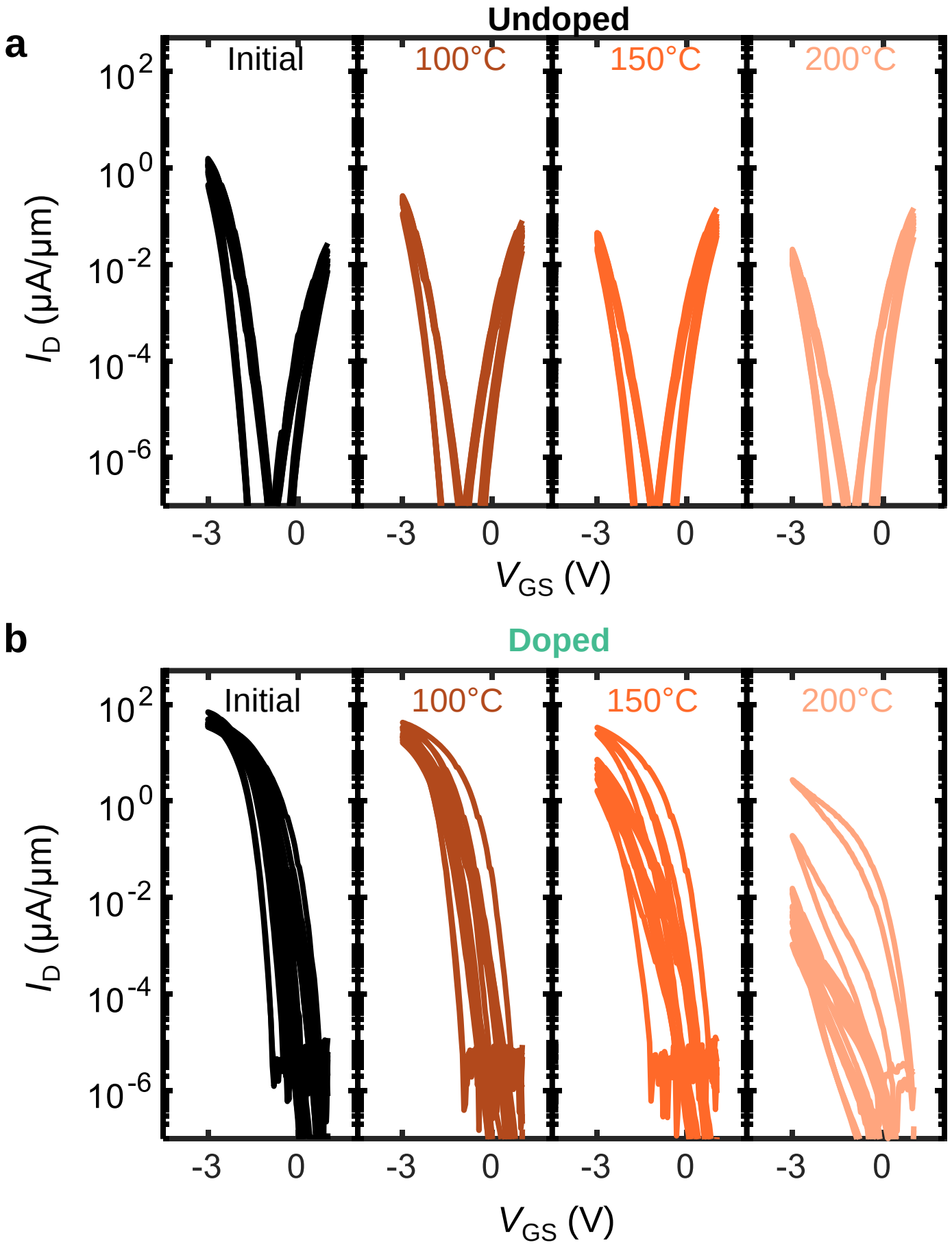

**Supplementary Fig. S8 | Thermal Stability of $L_{ch}$ = 1 μm Monolayer $WSe_2$ Devices. a,** Measured $I_D$ vs. $V_{GS}$ initially and after annealing in vacuum at 100, 150, and 200°C. 9 devices were measured and plotted. **b,** Measured $I_D$ vs. $V_{GS}$ initially and after annealing in vacuum at 100, 150, and 200°C for chloroform doped $WSe_2$. 14 devices were measured and plotted. For both plots, forward and backward sweeps are shown for all devices, and all devices demonstrate counterclockwise hysteresis.

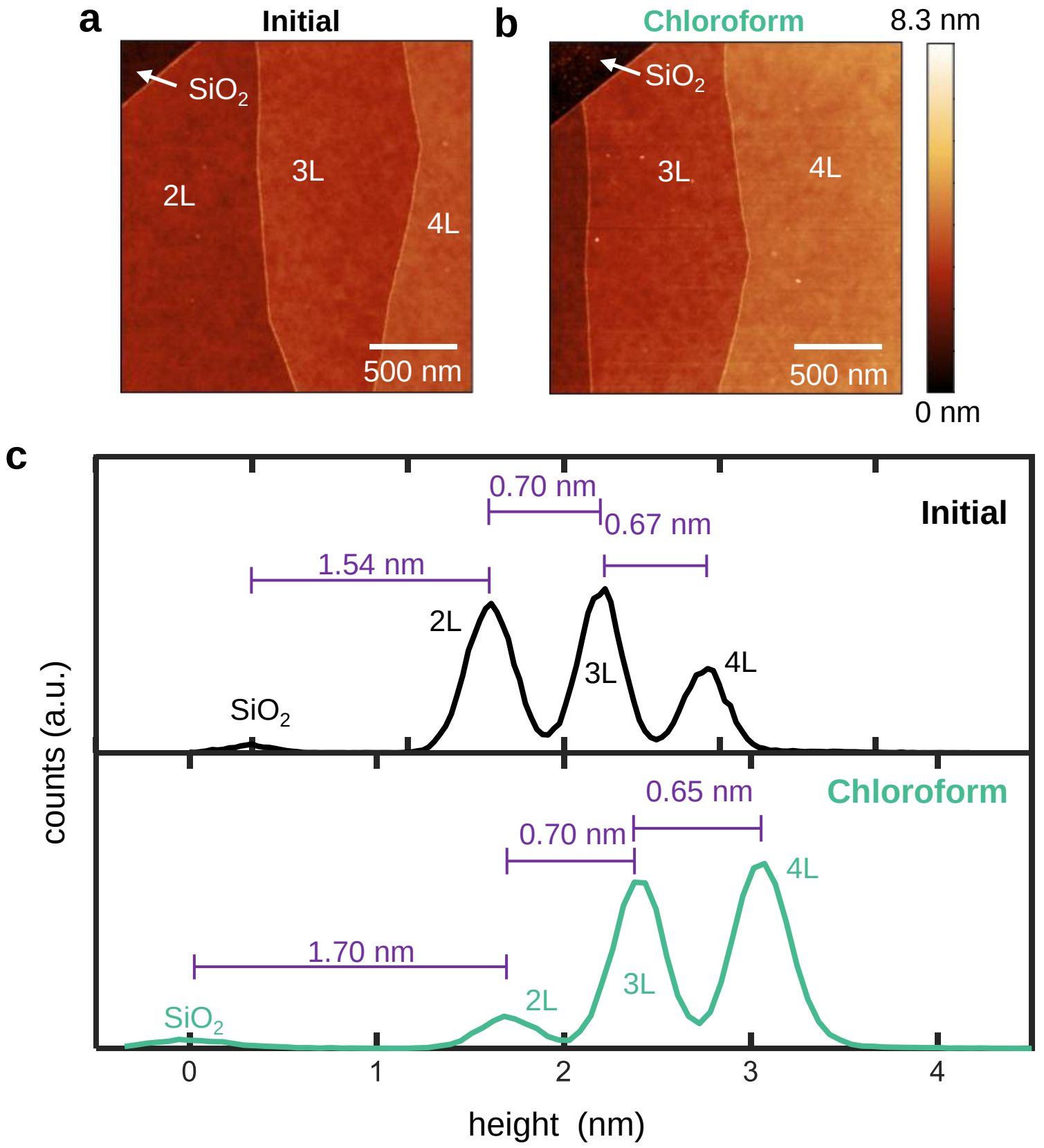

**Supplementary Fig. S9 | Atomic Force Microscopy (AFM) of Exfoliated $WSe_2$. a**, 2 × 2 μm$^2$ AFM scan of the 2-4L exfoliated $WSe_2$ flake in **Figure 5a** before doping. **b**, 2 × 2 μm$^2$ AFM scan of the 2-4L exfoliated $WSe_2$ flake in **Figure 5a** after chloroform doping. Some small particles can be identified on the flake after chloroform doping, similar to **Supplementary Fig. S2b**, but overall, there is minimal residue. **c**, Height distribution of the initial (top) and doped (bottom) AFM images in panel a and b, respectively. The peaks mark the height of the $SiO_2$, 2L, 3L, and 4L $WSe_2$ regions. There is no noticeable change in spacing between the $WSe_2$ layers, but the height difference between $SiO_2$ and 2L $WSe_2$ increases by > 0.15 nm. Gaussian curves were fit to the data to find the peak positions and peak separations.

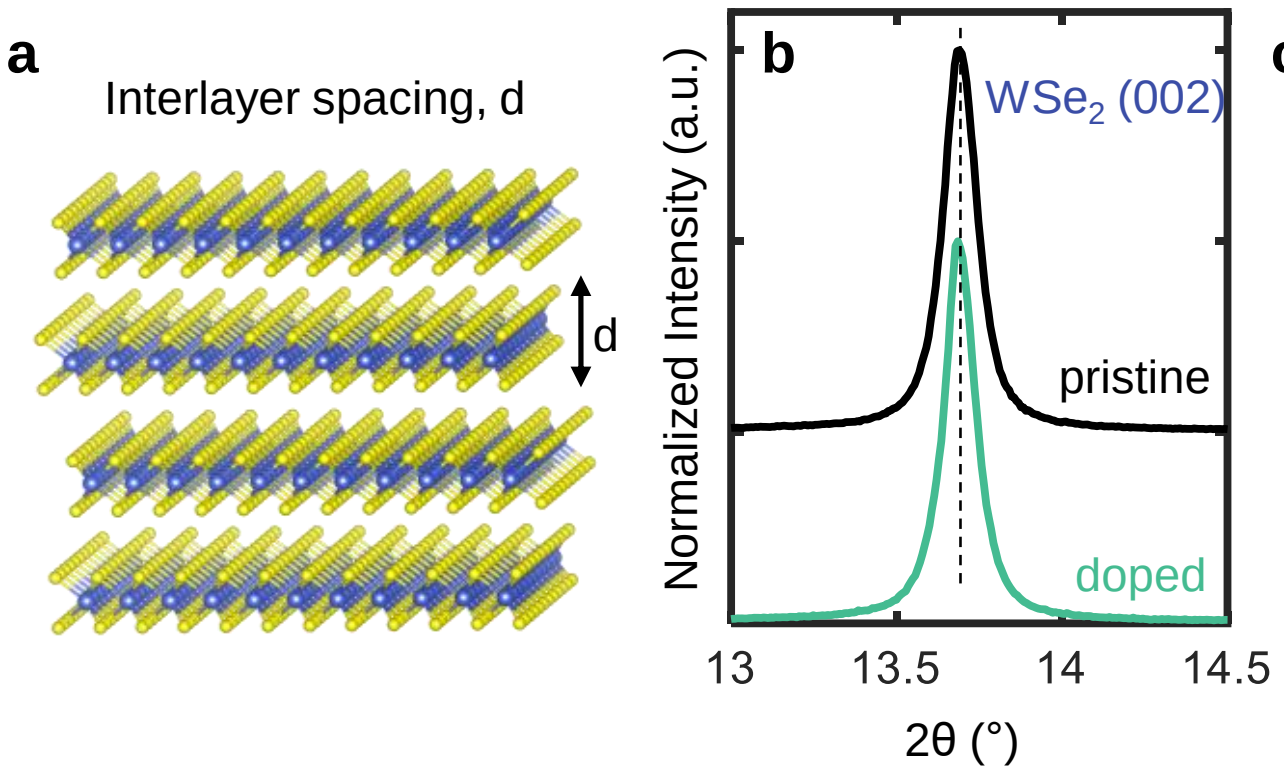

c

| | 2θ (°) | FWHM (°) | *d* (Å) |
|---|---|---|---|
| $WSe_2$ (pristine) | 13.693 | 0.12 | 6.461 |
| $WSe_2$ (doped) | 13.689 | 0.11 | 6.463 |

**Supplementary Fig. S10 | X-ray Diffraction (XRD) of Exfoliated Bulk $WSe_2$.** **a**, Schematic of multilayer $WSe_2$ with interlayer spacing, *d*. **b**, X-ray diffraction spectra of multilayer $WSe_2$ (002) peak with and without chloroform doping. **c**, Calculated interlayer spacing and full-width half-maximum (FWHM) for the peaks in panel b.

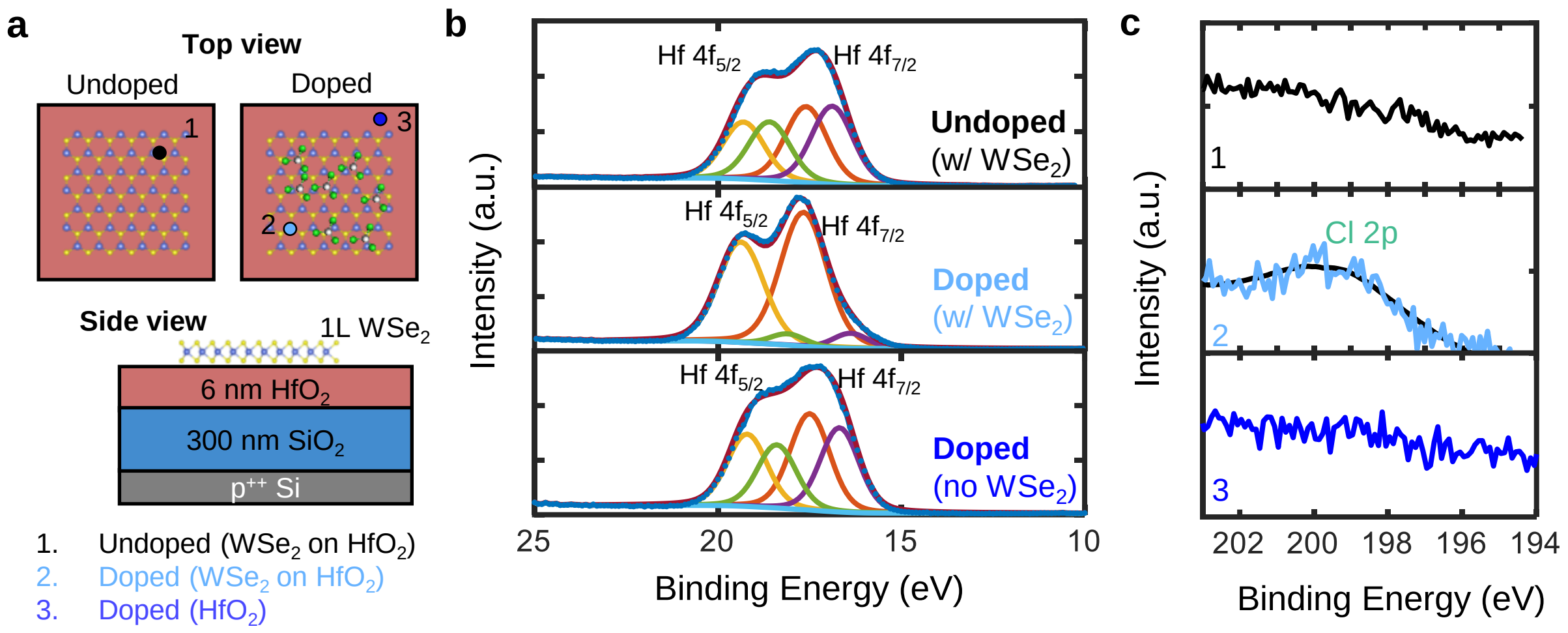

**Supplementary Fig. S11 | X-ray Photoelectron Spectroscopy (XPS) of Monolayer $WSe_2$ on $HfO_2$. a**, Representative schematic of 3 points probed by XPS: undoped $WSe_2$ on $HfO_2$, doped $WSe_2$ on $HfO_2$, doped $HfO_2$ sample with no $WSe_2$. $HfO_2$ was deposited by thermal atomic layer deposition, then underwent an $O_2$ plasma treatment prior to the $WSe_2$ transfer process. **b**, XPS spectra of Hf 4f peaks at the 3 points located in panel a. The Hf 4f peaks shift to higher binding energy after exposure to chloroform. This could indicate the existence of chloroform near the $HfO_2$ surface, which then passivate interface traps and decrease the subthreshold swing. **c**, XPS spectra of the Cl 2p peak at the 3 points shown in panel a, indicating that the Cl peak exists only in the doped $WSe_2$ region.

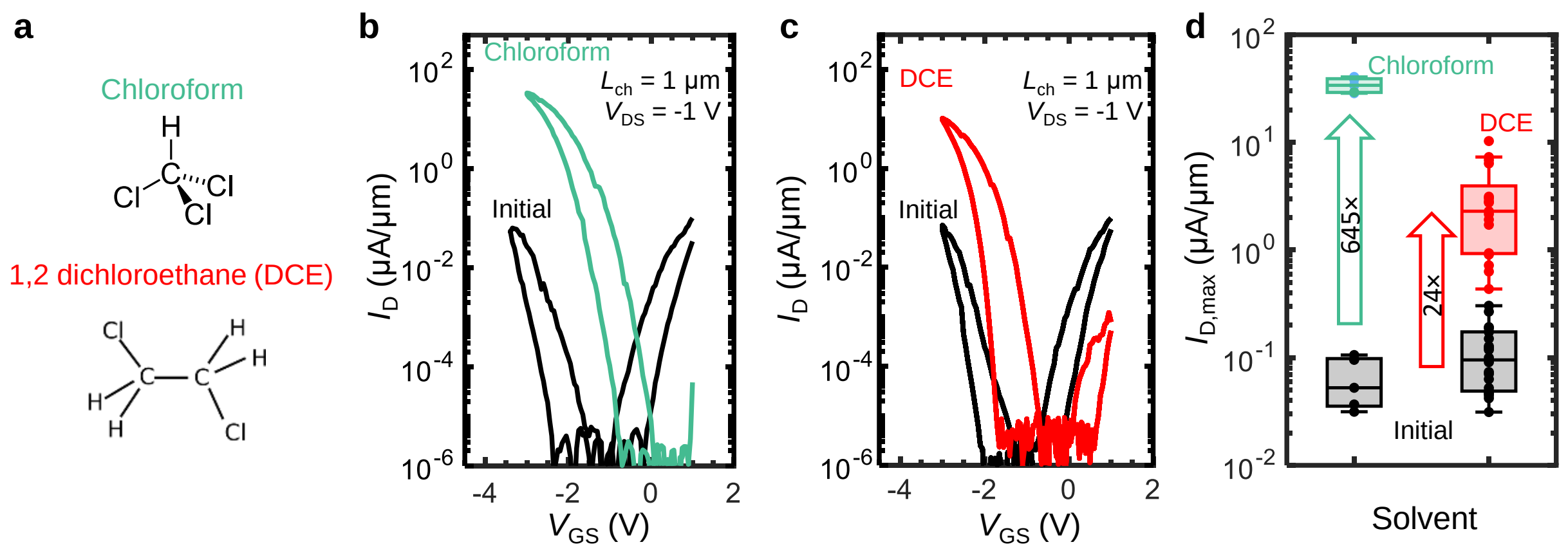

**Supplementary Fig. S12 | Effect of Dichloroethane (DCE) as a *p*-type Dopant in Comparison to Chloroform. a**, Diagram of a chloroform (top) and 1,2 dichloroethane (DCE) molecule (bottom). **b**, Measured $I_D$ vs. $V_{GS}$ curves of a monolayer $WSe_2$ device before and after doping with chloroform. Forward and backward sweeps are displayed with counter-clockwise hysteresis. **c** Measured $I_D$ vs. $V_{GS}$ curves of a monolayer $WSe_2$ device before and after doping with DCE. Forward and backward sweeps are displayed with counter-clockwise hysteresis. The devices were immersed overnight (> 8 hours) in dichloroethane (SIGMA-Aldrich, No. 34872) after device fabrication. **d**, Maximum drain-current ($I_{D,max}$) at $V_{GS}$ = -3.0 V before and after doping with chloroform and DCE. Chloroform doped devices show >10× greater $I_{D,max}$ compared to DCE devices. All devices shown here are with $L_{ch}$ = 1 μm. DCE increases the *hole* current of $WSe_2$ devices, with a median improvement in $I_{D,max}$ of ~24×, compared to the initial undoped $WSe_2$ devices. On the other hand, chloroform doping improves the hole current by ~ 645× on equivalent devices. The extra Cl atom in chloroform would increase the dipole moment of the dopant molecule, and thus could stabilize extra electrons, increasing the charge efficiency of chloroform. Additionally, the greater steric hinderance imposed by the larger DCE molecule suggests that it is much less likely to slip underneath the $WSe_2$.

**Table S1:** Benchmarking the Electrical Performance of Monolayer $WSe_2$

| Ref | Contact metal | Dopant | $L_{ch}$ (nm) | $I_{on}$ (μA/μm), $\|V_{DS}\|$ = 1 V | $I_{on}/I_{off}$ | $R_C$ (kΩ·μm)* | Stability (time, temperature) |
|---|---|---|---|---|---|---|---|
| 9 | Pd | - | 600 | 10 | $10^8$ | 127 | - |
| 10 | Ru | - | 200 | 50 | $2\times10^7$ | 2.7 | - |
| 11 | Ru | - | 100 | 100 | $10^8$ | - | - |
| 12 | Pt | - | 1500 | 7.6 | $2\times10^5$ | 229 | - |
| 13 | Pt/Au 10/80 nm | - | 700 | 108.1 | $2\times10^8$ | - | - |
| 14 | Ru/Au | - | 70 | 247 | $10^8$ | - | - |
| 15 | Pt/Au 20/60 nm | - | 200 | 15 | $10^7$ | - | - |
| 16 | $TiO_2$/Ru | - | 140 | 100 | $10^8$ | - | - |
| 17 | Ru | - | 50 | 92 | $10^8$ | - | - |
| 18 | Ti/Pd/Ni 1/30/30 nm | NO (175°C, 4 h) | 65 | 300 | $2\times10^6$ | 0.95 | 16 days |
| | | | 85 | 213 | | | |
| | | | 180 | 153 | | | |
| | | | 380 | 124 | | | |
| 19 | Sb/Pt 10/12 nm | 10 nm $MoO_x$ | 100 | 130 | - | 0.75 | - |
| | | - | 100 | 30 | $10^6$ | - | - |
| 20 | Sb/Pt 10/12 nm | $MoO_x$ | 100 | 170 | $10^7$ | - | - |
| | | - | 100 | 66 | $10^7$ | - | - |
| 21 | Pd | $WO_x$ | 500 | 82 | $10^7$ | ~1 | - |
| | | - | 1000 | 27.6 | $10^7$ | | - |
| 22 | Pd | 5 nm $MoO_x$ | 50 | 410 | $\sim10^6$ | 0.72 | - |
| | | - | 200 | 170 | $\sim10^7$ | | - |
| 23 | Ti/Pt/Au 0.5/30/30nm | 2× $WO_x$, NO | 55 | 550 | $\sim10^9$ | 0.875 | - |
| | | | 180 | 188 | | | - |
| | | | 380 | 105 | | | - |
| | | | 780 | 82 | | | - |
| | | - | 55 | 20 | $\sim10^8$ | - | - |
| 24 | Ti/Pt/Au 0.5/30/30nm | NO | 55 | 300 | $10^9$ | 0.875 | 24 days |
| 25 | Pd 25 nm | $HAuCl_4$ (50mM) | 5500 | 100 | 10 | - | - |
| | | $HAuCl_4$ (5mM) | 5500 | 8 | $10^7$ | - | - |
| 3 | Pd/Au 5/50 nm | $WO_xSe_y$ | 50 | 154 | $4\times10^7$ | 1.2 | - |
| 26 | few layer graphene | α-$RuCl_3$ | 500 | 31 | $10^9$ | 4.0<br>4.5 (at 10 K) | - |
| 27 | few layer graphene | α-$RuCl_3$ | - | - | - | 1.7 (at 300 mK, uniform doping), 20 (at 300 mK contact doping) | - |
| | Pt | α-$RuCl_3$ | - | - | - | ~ 36 (at 1.5 K) | - |
| **This Work** | Pd/Au 20/20 nm | $CHCl_3$ | 1000 | 119.2 | $10^{10}$ | 2.5<br>1.0 (at 10 K) | 243 days (8 months) 150 °C vacuum anneal |
| | | | 700 | 153 | $10^{10}$ | | |
| | | | 500 | 163.2 | $10^{10}$ | | |
| | | | 300 | 184.9 | $10^{10}$ | | |
| | | | 200 | 201.5 | $10^{10}$ | | |
| | | | 100 | 208.7 | $10^{10}$ | | |

*$R_C$ value given at room temperature if not otherwise stated

Blue text indicates *p*-type doping was implemented in the work

**Supplementary References:**